\documentclass[sensors,article,accept,moreauthors,pdftex,10pt,a4paper]{mdpi} 

\usepackage{pdfpages}
\usepackage{blindtext}
\makeatletter
\newcommand\xleftrightarrow[2][]{%
  \ext@arrow 9999{\longleftrightarrowfill@}{#1}{#2}}
\newcommand\longleftrightarrowfill@{%
  \arrowfill@\leftarrow\relbar\rightarrow}
\makeatother
\usepackage{amsfonts}
\usepackage{array}
\newcolumntype{C}[1]{>{\centering\let\newline\\\arraybackslash\hspace{0pt}}m{#1}}
\firstpage{1} 
\usepackage{makecell}
\usepackage{multirow}
\usepackage[utf8]{inputenc}
\usepackage{pifont}
\usepackage{newunicodechar}
\usepackage{enumitem,microtype}
\graphicspath{{./Figures/}}
\usepackage{scalerel}
\usepackage{wrapfig}

\articlenumber{979}
\doinum{10.3390/s17050979}
\pubvolume{17}
\pubyear{2017}
\copyrightyear{2017}
\externaleditor{}
\history{DOI: 10.3390/s17050979}

 \theoremstyle{mdpi}
 \newcounter{thm}
 \newcounter{ex}
 \newcounter{re}
\Title{An Authentication Protocol for Future Sensor~Networks}

\Author{Muhammad Bilal * and Shin-Gak Kang }
\AuthorNames{Muhammad Bilal and Shin-Gak Kang}

\address[1]{%
Electronics and Telecommunications Research Institute, University of Science and Technology, 218,~Gajeong-ro, yuseong-gu, Daejeon 34129,
Korea; sgkang@etri.re.kr (S.-G.K.)\\}

\corres{\hspace{-.75em}Correspondence: mbilal@etri.re.kr; Tel.: +82-42-860-6424}

\abstract{Authentication is one of the essential security services in Wireless Sensor Networks (WSNs) for ensuring secure data sessions. Sensor node authentication ensures the confidentiality and validity of data collected by the sensor node, whereas user authentication guarantees that only legitimate users can access the sensor data. In a mobile WSN, sensor and user nodes move across the network and exchange data with multiple nodes, thus experiencing the authentication process multiple times. The integration of WSNs with Internet of Things (IoT) brings forth a new kind of WSN architecture along with stricter security requirements; for instance, a sensor node or a user node may need to establish multiple concurrent secure data sessions. With concurrent data sessions, the frequency of the re-authentication process increases in proportion to the number of concurrent connections. Moreover, to establish multiple data sessions, it is essential that a protocol participant have the capability of running multiple instances of the protocol run, which makes the security issue even more challenging. The currently available authentication protocols were designed for the autonomous WSN and do not account for the above requirements. Hence, ensuring a lightweight and efficient authentication protocol has become more crucial. In this paper, we present a novel, lightweight and efficient key exchange and authentication protocol suite called the Secure Mobile Sensor Network (SMSN) Authentication Protocol. In the SMSN a mobile node goes through an initial authentication procedure and receives a re-authentication ticket from the base station. Later a mobile node can use this re-authentication ticket when establishing multiple data exchange sessions and/or when moving across the network. This scheme reduces the communication and computational complexity of the authentication process. We proved the strength of our protocol with rigorous security analysis (including formal analysis using the BAN-logic) and simulated the SMSN and previously proposed schemes in an automated protocol verifier tool. Finally, we compared the computational complexity and communication cost against well-known authentication protocols. }


\begin{document}



\section{Introduction}
Wireless sensor networks (WSNs) consist of a vast number of distributed sensor nodes. Each~sensor node is an autonomous system that monitors and collects data from the surrounding environment. In wireless sensor networks, the sensor nodes have limited computational power and communication capabilities; hence, most of the conventional cryptographic mechanisms and security protocols are not suitable for resource limited WSNs. For instance, very efficient public key algorithms, such as ECC
 \cite{1}, need a fraction of a second to execute the encryption/decryption procedures, while a symmetric key algorithm, such as RC5 \cite{2}, needs only a fraction of a millisecond to perform encryption and decryption procedures \cite{3,4,5}. In a sensor network where devices have limited resources, the asymmetric cryptographic functions must be used wisely; for instance, use the asymmetric cryptography when an authentication responder already verified the initiator and wants to share secret information. If the responder does not authenticate the initiator and the initiator uses the asymmetric cryptography to exchange a message, the network becomes vulnerable to DOS
  attacks \cite{5,6}. The WSN related contraints mentioned above are known to the research community. The currently available authentication protocols are designed for the autonomous WSN from the perspective of the above constraints.
Moreover, in the near future in the realization of the vision of emerging technologies such as IoT, D2D,
 smart home and smart cities, WSNs will provide an invaluable service by acting as a virtual layer between the physical world and the computational devices \cite{7,8,9}. However, integration of WSNs with IoT will bring forth a new kind of WSN architecture and stricter security requirements; for instance, in a smart hospital (as shown in Figure~\ref{fig1}b) a sensor node or a user node may require the establishment of multiple concurrent secure data sessions. To establish a secure data session, authentication is the first step. In a dynamic, mobile WSN environment, where sensors and user nodes can establish multiple concurrent connections, a node moving across the network undergoes the authentication check multiple times and the frequency of the re-authentication process increases in proportion to the number of concurrent connections. Moreover, to establish multiple data sessions, it is essential that a protocol participant has the capability of running multiple instances of the protocol run, which makes the security issue even more challenging. Thus, it is essential to adopt a secure yet lightweight authentication procedure that especially reduces the computational time and communication at the mobile sensor node. The currently available authentication protocols were designed for the autonomous WSN and do not account for these new emerging challenges. 
This work presents a novel authentication protocol suite called the Secure Mobile Sensor Network (SMSN) Authentication Protocol. The SMSN protocol suite consists of six protocols: three protocols deal with mobile sensor node authentication with sink nodes and the other three deal with user node activation and authentication with the base station, sink nodes, and sensor nodes. In the SMSN, mobile sensors and user nodes can join and leave the system dynamically and can establish secure multiple concurrent connections. After the initial authentication, a mobile sensor or a user node can move across the network and get re-authenticated by a simple ticket-based re-authentication protocol; for instance, a user node can establish concurrent connections with multiple sink and sensor nodes using a re-authentication ticket issued during the initial-authentication protocol run. To~establish multiple connections, a node is allowed to run multiple instances of the protocol; consequently, we introduce extra design requirements to meet the goals of a secure authentication protocol. In this paper, we also present an efficient and lightweight key generation and distribution mechanism. In the key generation protocol, a commitment key is generated by a group of participants (the base station and sink nodes) using an irreversible function; the key agreement and key retrieval protocol are the same as that employed in \cite{10}. The commitment key is further applied to drive multiple time-based encryption keys, for example, the ticket encryption key and session key between the sink and user/sensor are derived from the commitment key. The time dimension in the protocol increases the security of the protocol; although the group members do not need to be tightly or loosely time synchronized.
To~determine the security of the protocol in this study, we performed a rigorous security analysis and also simulated the SMSN and previously proposed schemes \cite{16,17,18,19,20,21}
 in an automated security protocol analysis tool called Scyther \cite{11,12}, which is a powerful state-of-art tool that finds attacks for defined protocol properties. We observed that our authentication protocol is secure, and it achieves all the objectives of an authentication protocol, which are defined as protocol claims in Section 4.3; for a detailed description of protocol claims, please refer to \cite{13,14}. We also compared the efficiency of the SMSN in terms of computational time and communication complexities as discussed in \cite{16,17,18,19,20,21}. The~remainder of this paper is organized as follows. Section 2 presents the related work. In~Section 3, we present a brief system overview and problem statement. Section 4 describes the proposed scheme with a detailed discussion. In Section 5, we assess the strength of our scheme against the known attacks. Section 6 presents an efficiency analysis that compares a few interesting schemes with our scheme. Finally, we provide concluding remarks in Section 7. 
\section{Related Work}
Typically, WSNs are comprised of distributed devices with limited resources. Most of the conventional cryptographic mechanisms and security protocols are computationally expensive and are not suitable for resource-limited WSNs. In the recent past, the research community proposed several authentication protocols \cite{16,17,18,19,20,21,15,22,23,24,25,26,27} that provide security in a WSN environment. Since the sensor nodes have low computational time, storage and communication capabilities, it is essential to design an efficient and lightweight yet secure authentication mechanism. From the point of view of computational and communication complexity, the authentication procedure in a wireless network with a mobile sensor and user node is an expensive task. A node moving across the network undergoes multiple authentication checks. Thus, it is essential to adopt a secure yet lightweight authentication procedure that especially reduces the computational and communication resources at the mobile sensor node. In \cite{28} the authors discussed various anomaly detection techniques for flat and hierarchical wireless sensor networks, but detection techniques are not sufficient for several security threats. However, for a secure system together with detection methods, prevention techniques such as authentication is also vital; various authentication protocols for WSNs were proposed in  \cite{16,17,18,19,20,21,15,22,23,24,25,26,27}.
In 2006, Wong~\cite{15} proposed a user authentication scheme for a dynamic WSN. The scheme is password based and employs lightweight cryptographic hash and XOR operations. Later on, Tesng~et~al.~\cite{16} and Das \cite{22} identified that the Wong \cite{15} scheme had various weaknesses and was vulnerable to replay attacks and forgery attacks, and the user password is known to the sensor node and can be revealed by any sensor node. Tesng~et~al.~\cite{16} proposed an improved version to mitigate the weaknesses that posed security threats in the Wong \cite{15} scheme. The scheme of Tesng~et~al.~\cite{16} is also a password-based scheme, but the password is not revealed to the sensor nodes, and they also introduced a new phase of password change. Nevertheless, the scheme of Tesng~et~al.~\cite{16} is weak against replay attacks, impersonation attacks and forgery attacks \cite{29}. Moreover, the scheme does not provide a mutual authentication between the gateway (GW) and sensor node (SN).
In 2009, Das~\cite{22} proposed a two-factor user authentication scheme in which legitimate users can register and log in to the remotely deployed sensor nodes to access the collected data. From the point of view of computational and communication complexity, the scheme is reasonably efficient. The author claimed the scheme was secure against various kinds of attacks. However, later work \cite{30,31,32} suggested that the scheme was vulnerable to different types of attacks, including impersonation, password guessing, insider, and parallel session attacks. Moreover, it did not provide the mutual authentication between the GW and sensor nodes.
In 2010, Yoo~et~al.~\cite{17} proposed a user authentication scheme for WSNs and analyzed the protocol using BAN logic \cite{33}. However, the BAN logic provided a foundation for the formal analysis of security protocols, but in the case of authentication, various attacks could slip through the BAN logic~\cite{34,35}. The scheme of Yoo~et~al.~provided mutual authentication between the GW and the user and established a session key between the GW and SN. The authors claimed that the scheme was safe against insider attacks, impersonation attacks, and parallel session attacks. However, in \cite{27} the authors provided a detailed analysis of the scheme of Yoo~et~al.~and proved that the scheme is susceptible to various attacks, including insider attacks, impersonation attacks, parallel session attacks, password guessing attacks, fake registration attacks, and DOS attacks. 
Kumar~et~al.~\cite{36} proposed an authentication protocol for WSNs and claimed it could satisfy all the security requirements of the WSN; however, He et al.~\cite{37} proved that the scheme was weak against insider attacks and offline password guessing attacks and could not provide user anonymity. Kumar proposed another enhanced scheme in \cite{18} and once again claimed that the scheme could withstand most of the known attacks and provide user privacy. However, with the Scyther implementation, given in Section 4.6, we found that the improved scheme presented in~\cite{18} was still vulnerable to insider attacks, parallel session attacks, and impersonation attacks. 
Farash~et~al.~\cite{20} proposed a key agreement and authentication protocol for WSNs in the Internet of Things (IoT) environment. The scheme was well designed and provided security against several well-known attacks. The author proved the strength of the protocol with BAN logic and further confirmed the theoretical analysis results by implementing the protocol in the AVISPA~\cite{38} tool. However, similar to other schemes discussed above, \scalebox{.95}[1.0]{Farash~et~al.~\cite{20}} did not consider the requirement of concurrent sessions, which are more likely to occur in an IoT environment. Moreover, our implementation of the scheme of Farash~et~al.~\cite{20} in Scyther revealed that the GW was vulnerable against insider attacks and impersonation attacks. The~scheme was insecure in the presence of an intruder as discussed in Section 4.6, which assumed that the initial knowledge set of intruders included the identities of all sensor and user nodes. With known identities, an intruder can impersonate the user node and deceive the GW to falsify the authentication properties. Similarly, the Scyther implementation of the recently published work of Y. Lu~et~al.~\cite{21} and Quan~et~al.~\cite{19} revealed that at least two protocol participants falsified the authentication properties. The majority of the schemes discussed above provided the GW and user authentication schemes; however, Farash~et~al.~\cite{20} and Y. Lu~et~al.~\cite{21} also considered the GW and sensor node authentication, while Farash~et~al.~\cite{20} also allowed the user to access the data from sensor nodes directly. 
Li~et~al.~ \cite{39} proposed an authentication protocol for sensor and user nodes. However, the proposed scheme requires time synchronization among protocol participants, and also employs the asymmetric elliptic curve cryptography, which is not a good design choice for resource limited WSN applications. For~instance, very efficient public key algorithms, such as ECC \cite{1}, need a fraction of a second to execute the encryption/decryption procedures, while a symmetric key algorithm , such as RC5 \cite{2}, needs only a fraction of a millisecond to perform  encryption and decryption procedures \cite{3,4,5}.
Unlike all the schemes discussed above, our proposed scheme, the SMSN authentication protocol suite, allows the sensor and user nodes to establish multiple concurrent connections with different sensor and sink nodes, which makes our scheme suitable for deploying it in the future to support IoT and related emerging technologies. Moreover, the SMSN authentication protocol suite provides several kinds of mutual authentications; for instance, after the initial authentication, a sensor or the user node receives an authentication ticket issued by the base station. The ticket can be further used for sensor-sink, user-sink, and user-sensor mutual authentication. The SMSN protocol suite consists of six protocols: three protocols deal with mobile sensor node authentication with sink nodes, and the other three deal with user node activation and authentication with the base station, sink nodes, and sensor nodes.

\section{System Overview and Problem Statement}
A typical WSN consists of the base station ($B$S), sink node ($S$), sensor node ($N$) and user node ($U_{i})$. We 
assume that the $BS$ knows the public keys of the sink ($S$), sensor ($N$) and user ($U)$ nodes.

\subsection{System Architecture}
An IoT smart service provider deploys the WSN with various base stations 
connected to the internet through a service center ($SC)$. In the 
WSN, each $BS_{j}$ forms a group $G_{j} $ consisting of neighbor base 
stations and associated sink nodes. All group members share a symmetric 
group key $K_{G}^{j} $, which is controlled by the group master $
BS_{j}$ using a group key agreement protocol such as discussed in \cite{40}. Furthermore, the base station can access and download the 
profile of the mobile Sensor ($N$) and User ($U) $ nodes. Each 
profile has a unique secret number $n_{s}^{i}$ ; besides being know to 
the base station, this secret number is also known to the corresponding 
Sensor ($N$) and User ($U)$ nodes. The profile and unique secret 
number $n_{s}^{i}$ of all legitimate users and sensor nodes are 
accessible to the base station through the service center. 

An example of the overall system architecture is depicted in Figure 
1a. $BS_{1}$ creates a group consisting of neighbor $BS_{0}$ and 
$BS_{2}$ and associated sinks $ (S_{1},S_{3})$. $BS_{1} $ generates 
a group key $K_{G}^{1}$ and shares it among all group members. The 
profiles and the associated unique secret number of $U_{1}, N_{1}, $
and $N_{2}$ are accessible to $BS_{0}, BS_{1}, $ and $BS_{2}$ via 
the service center. An application scenario is given in Figure 1b. In 
a smart hospital, the health status of a hospitalized patient is 
continuously monitored by the several sensors mounted over the patient's 
body. For the secrecy of patients' health data it is essential that only 
authorized users be able to access the data. When a patient takes a walk 
in the hospital, all the mounted sensors continuously transfer the 
different sets of data to the nearby sinks. With the SMSN, the sensors 
do not need to follow the full authentication procedure for each data 
session; instead, the session continues by sharing a simple 
re-authentication ticket that provides the sensor-sink mutual 
authentication. Similarly, when an authorized doctor or nursing staff 
visits patients, he or she can access the sensor data in real time via a 
user device by establishing multiple data sessions with various sinks 
and sensor nodes. While he moves from patient to patient, his device 
does not need to follow a complete authentication procedure for each 
data session; rather, a simple re-authentication ticket can be used for 
re-authentication. From outside of the hospital, an authorized person 
can log into the hospital system and access the necessary data from sink 
and base stations by using the same re-authentication ticket. 
\begin{figure}[H]
\centering
\includegraphics[width=14 cm]{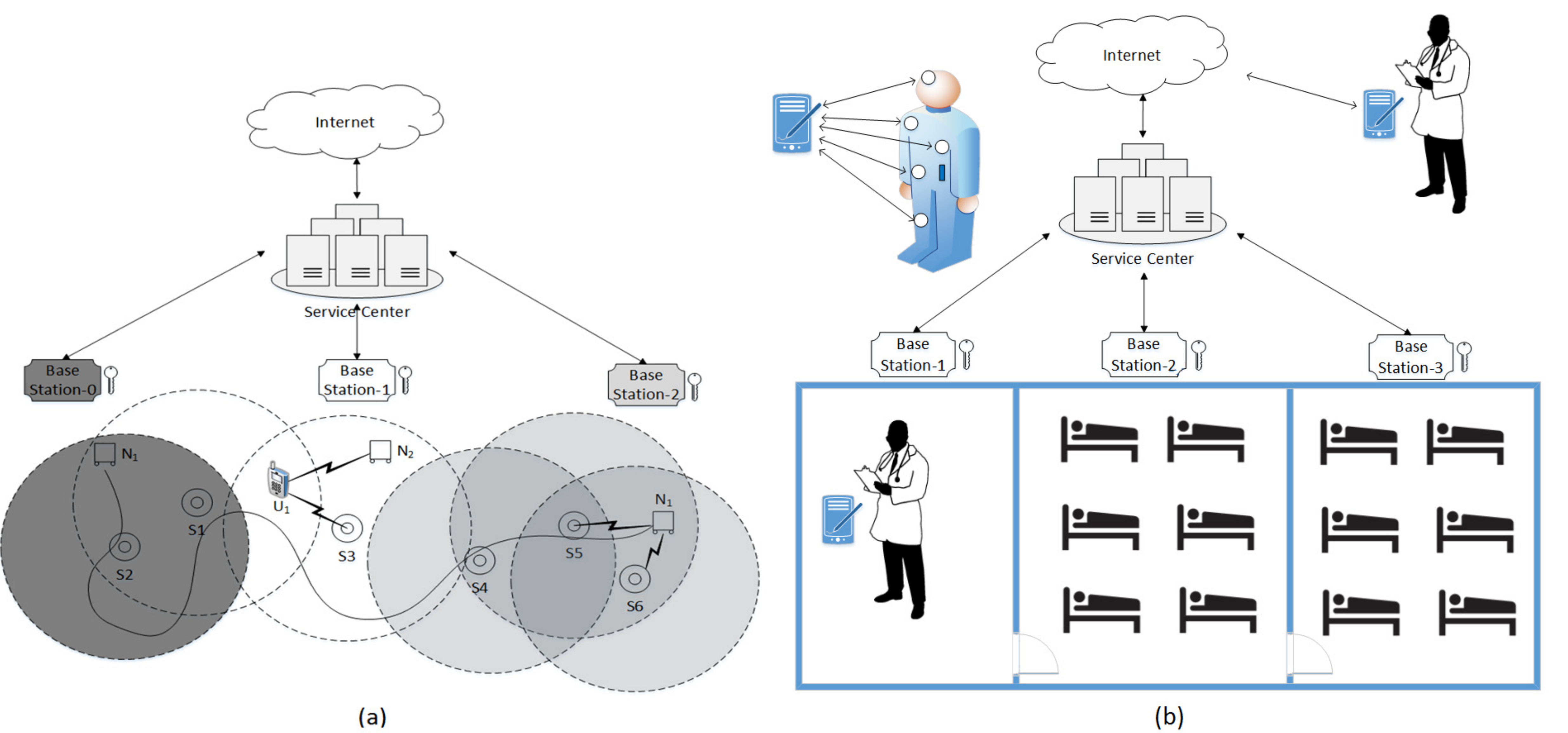}
\caption{({\bf a}) An example scenario of a Wireless Sensor Network (WSN). Sensor node $N_1$is initially authenticated by $BS_1$; while moving across the network it is re-authenticated by the re-authentication ticket. At the final destination $N_1$  shares data with multiple sink nodes while user node $U_1$ collects data from sensor node $N_2$ and sink node $S_3$. ({\bf b}) An application scenario of  Secure Mobile Sensor Network (SMSN) in a smart hospital. Authorised doctors and nursing staff can access the data from sensor nodes using a user device.}\label{fig1}
\end{figure}   
\subsection{Problem Statement }
 As described earlier, in IoT and other related emerging technologies, a 
mobile sensor or user node may need to exchange data with multiple nodes 
and so will experience the authentication process multiple times. With 
multiple concurrent connections, the authentication process becomes even 
more expensive while a node moves across the network. As shown in Figure 
\ref{fig1}a, sensor node $N_{1}$ is communicating with sink nodes $S_{5}$ 
and $S_{6}$ simultaneously. Likewise, user node $U_{1}$ is 
communicating with sink node $S_{3}$ and sensor node $N_{2}$ 
simultaneously. In such an application scenario, when a sensor or the 
user node moves across the network the frequency of the 
re-authentication process increases in proportion to the number of 
concurrent connections. Moreover, to establish multiple data sessions, 
it is essential that a protocol participant run multiple instances of 
the protocol run, which makes the security issue even more challenging. 
To perform multiple parallel re-authentications, it is evident that the 
protocol participants run multiple instances of the protocol. With the 
multiple protocol runs, the assurance of security of the protocol 
becomes more challenging. Therefore, for seamless services, lightweight 
yet secure re-authentication is vital. 

According to our knowledge, we are the first to propose an 
authentication protocol for concurrent secure connections. To perform 
multiple parallel re-authentications, the protocol participants must run 
multiple instances of the protocol. With multiple protocol runs, the 
assurance of security of the protocol becomes more challenging. We 
developed our scheme, the SMSN, considering the following constraints: 
(1) the communication channels are insecure; (2) an intruder with the 
capabilities as described in Section-5-C is present in the network to 
launch various attacks; (3) due to the requirements for a WSN deployed in 
an IoT environment, the protocol participants are allowed to run 
multiple instances of the protocol; and (4) user and sensor nodes can 
dynamiclly leave and join the network and can move across the network. 

\subsection{Notations}
\begin{itemize}[leftmargin=*,labelsep=4mm]
\item $BS_{j}$ = The $j$th base station 
\item $S_{j}$ = The $j$th sink 
\item $N_{i}$ = The $i$th sensor node
\item $U_{i}$ = The $i$th user node
\item $A\sim B$ = A is associated with B such that B is controlling 
authority 
\item $G_{j}$ = Group of all associated entities of $j$th $BS$
\item $G_{j}^{o}$ = Group of all non-associated entities of $j$th
BS who knows the $K_{G}^{j}$
\item $K_{i}^{j}$ = The time-based key generated at $i$th interval by 
$BS_{j}$ 
\item $K_{G}^{j}$ = A group key generated by $BS_{j}$
\item $K_{S}^{i}$ = $ N\to S$ session key generated at $i$th interval
\item $K_{PS}^{i}$ = $N\to S$  private session key 
\item $K_{TS}^{i}$ = Temporary session key
\item $E_{B}^{i}(m)$ = Encryption of 'm', using key $K_{B}^{i}$ 
\item $V_{i}$ = The index value for interval i
\item $T_{k}$ = The $k$th Ticket 
\item $Z(A)$ = An intruder $Z$ mimicking the entity $A$
\item $n_{i}$ = $i$th nonce in a message exchange
\end{itemize}

\section{Proposed Scheme }
The SMSN protocol suite consists of two protocol suites, the Keying 
Protocol suite and an Authentication Protocol suite. The Keying Protocol 
suit further comprises a key agreement protocol, a key retrieval 
protocol (which is the same as the one employed in \cite{10}), and a key 
management protocol; likewise, the authentication protocol further 
comprises six protocols, three dealing with mobile sensor node 
authentication with sink nodes and the other three dealing with user 
node activation and authentication with a base station, sink nodes, and 
sensor nodes in different scenarios. In subsequent sections, the SMSN 
protocol suite is described in detail.
\subsection{Keying Protocol Suite }
In key generation protocol a 'commitment key' is generated by group 
participants (the base station and sink nodes) using an irreversible 
function similar to that as used in \cite{10,41}. The 'commitment key' is 
further used to drive multiple time-based keys; for instance, the ticket 
encryption key and session key between the sink and user/sensor are 
derived from the 'commitment key'.
\subsubsection{Key Agreement Protocol }
\begin{figure}[H]
\centering
\includegraphics[width=14 cm]{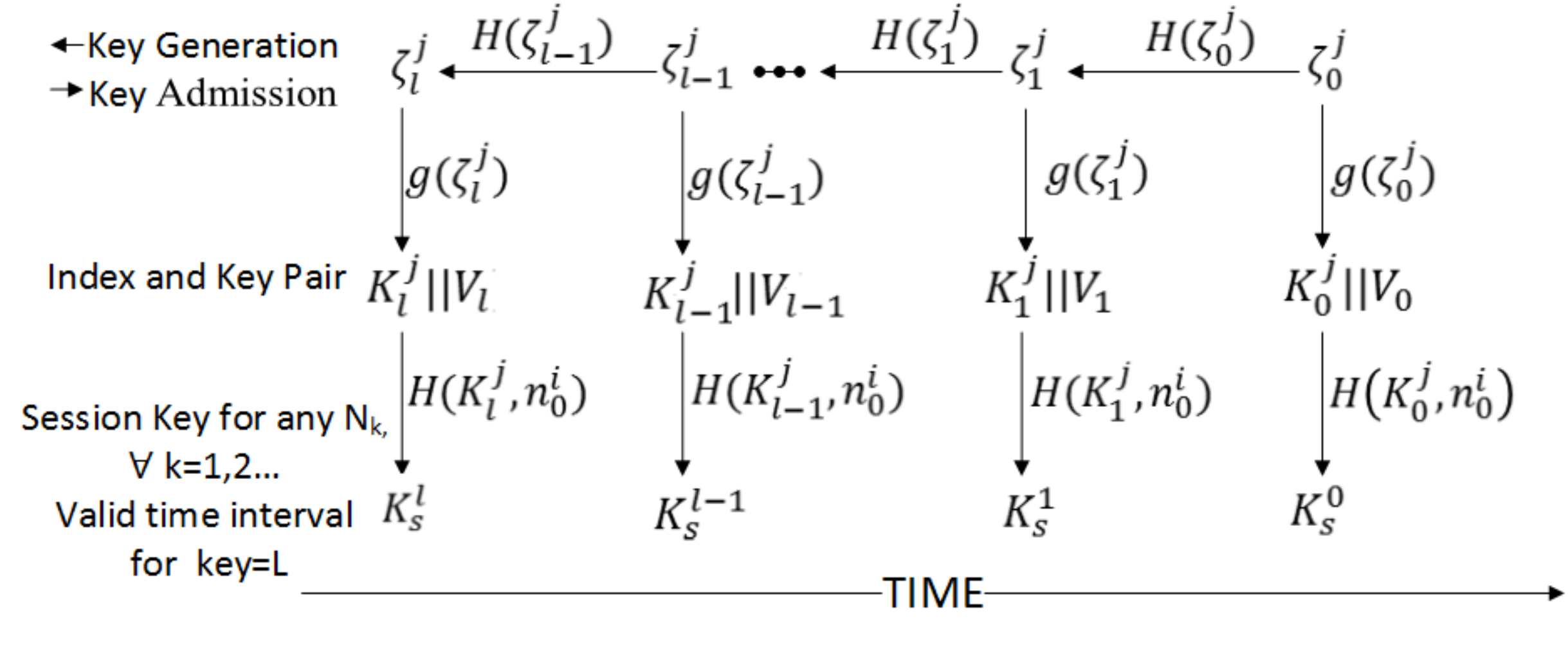}
\caption{Time-based keys generation and admission with reference to time passage.}\label{fig2}
\end{figure}

The key generation and distribution mechanism is shown in Figure \ref{fig2} and consists of the following 
steps: (1) After every time interval $ T_{d}$, $ BS_{j}$ broadcasts the key generation 
information ($
Key_{MSG}$) to all members of $ G_{j}.$ (2) Using the key generation 
information ($KEY_{MSG})$, all members generate a Commitment Key 
Generator $ \zeta _{0}^{j}=H(T_{d},N_{0}^{j})$. (3) All members of  $
G_{j}$ can now generate a ``\,Chain of Key Generators'' of length $L$ 
by using an irreversible one-way function: $ {H(\zeta _{0}^{j})=\zeta 
_{1}^{j},H(\zeta _{1}^{j})=\zeta _{2}^{j}\ldots H(\zeta 
_{l-1}^{j})=\zeta _{l}^{j}}$; i.e., $H(\zeta _{k}^{j})^{i}=\zeta 
_{k+i}^{j}.$ (4) Each generator ($\zeta )$ in the chain is used by 
function $g $at specific intervals to derive indexes and a ticket 
encryption key pair. For instance, at interval $k$ for any sensor or 
user node $i,$ which is requesting the $BS_{j}$ to join the network, 
the function $g(\zeta _{k}^{j})=H(\zeta 
_{k}^{j},H(n_{0}^{i}))||H(k) $ generates ticket encryption key $
K_{k}^{j}$ and index $V_{k}$ value. In function $g $ the value $
n_{0}^{i}$ is a secret nonce sent by node $i$ in a network join 
message. (5) Furthermore, $BS_{j}$ issues a session key based upon the 
ticket encryption key $K_{k}^{j}$ and secret nonce $n_{0}^{i}$, 
i.e., $K_{S}^{i}=H(K_{k}^{j},n_{S}^{i})$ where $n_{S}^i$ is the secret random number assigned to legitimate sensor and user node. 

All the symmetric keys generated in the above discussion have a size of 
256 bits (32 bytes); hence, in the subsequent section of authentication 
protocols any symmetric encryption supporting the 256-bit key can be 
used, e.g., RC5/6 \cite{2}; Rijndael \cite{42}, Twofish \cite{43}, MARS 
\cite{44}, and Blowfish \cite{45}. 

\subsubsection{The Key Retrieval Protocol }
After initial authentication, $BS_{j}$ issues a ticket to the 
requesting node. The ticket consists of two parts: (1) The first half 
consists of the sensor node identity (id) $N_{i}$, the session key $K_{s}^{i}$
, secret nonce $n_{0}^{i}$, and the profile. This part of the ticket 
is encrypted with time-based key $K_{l}^{i}$. (2) The second half 
consists of sensor node id $N_{i}$, the hash of group id $H(G^{j})$
, and the required information to retrieve the time-based key $
K_{l}^{i}$. The hash of group id $H(G^{j})$ is an optional field 
used only if sink nodes can join multiple base stations; in that case, 
it is used to identify the group and to select the correct keychain. In~the key retrieval, information depends on the selected mode and can be 
the scrambled index value $V$, index vector $V_{Hash}$, or index 
value $ i$; the modes of a ticket are explained below.

\textbf{Mode-01:}

In mode-01, the ticket retrieval information comprises the index value  
$(V_{i})$, requesting node id $N_{k}$, and the hash of the user's 
private key. The ticket verifier searches the appended index value $
(V_{i}) $ within its generated vector $(V)$. A search hit at the $
ith$ place means that key $K_{i}^{j}$ can decrypt the ticket.
\begin{equation}
\vspace{+6pt}
T_{k}=E_{i}^{j}(N_{i}||K_{s}^{i}||n_{0}^{i}|\vert 
Prof||E_{G}^{j}(N_{i}||V_{i}||H(G^{j})) \tag{Ticket formate for mode-01}
\end{equation}

\textbf{Mode-02: }

If the group members are in a fully secure environment, the SMSN employs 
a simple ticket retrieval strategy. Instead of using a scrambled index 
value $V $ in mode-02 in the ticket retrieval, the information contains 
the interval value. In this mode, the ticket verifier does not need to 
run a search~algorithm. 
\begin{equation}
\vspace{+6pt}
T_{k}=E_{i}^{j}(N_{i}||K_{s}^{i}||n_{0}^{i}|\vert 
Prof||E_{G}^{j}(N_{i}||i||H(G^{j}))
\tag{Ticket formate for mode-02}
\end{equation}

\newpage
\textbf{Mode-03: }

\begin{figure}[H]
\centering
\includegraphics[width=4.5 cm]{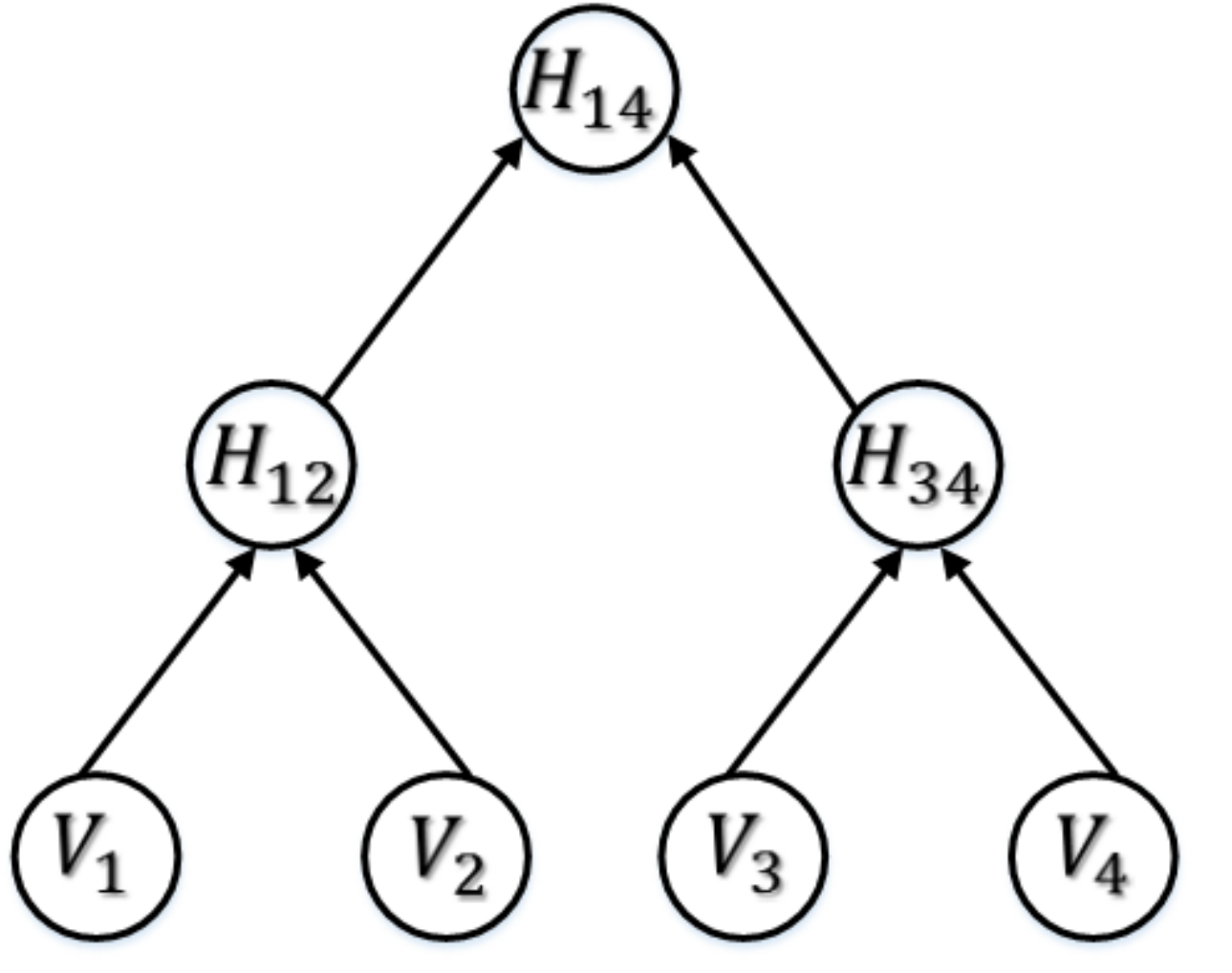}
\caption{Example of Binary Hash Tree generated with index vector $V={v_1,v_2,v_3,v_4}$.}\label{fig3}
\end{figure}

In mode-03, the ticket retrieval information is the same as in mode-01 
except it includes a hash vector of size $\log _{2}\vert V\vert -2$. 
These hash values are carefully chosen nodes of the binary hash tree, 
which is generated such that the leaf nodes are indexed vector $(V)$ 
values as shown in Figure \ref{fig3}. 
\begin{equation}
T_{k}=E_{i}^{j}(N_{i}||K_{s}^{i}||n_{0}^{i}|\vert 
Prof||E_{G}^{j}(N_{i}||V_{Hash}||H(G^{j}))
\tag{Ticket formate for mode-03}
\end{equation}

\textit{Search Algorithm:}\ \ \ \ 

\begin{itemize}[leftmargin=2.7em,labelsep=5mm]
\item Start from the root and move down 
\item Ignore appended values and follow the path of the reconstructed 
node
\item Continue until level $\log _{2}\vert V\vert -1$ is reached
\item At level $\log _{2}\vert V\vert $, select the appended value 
that is the index value.
\end{itemize}

Mode-03 is suitable if the Chain of the Key Generator is very long. The 
tree ``root node'' is included in $T_{k}$ (optionally), which ensures 
that a trusted group member generated $ T_{k}$.

\subsubsection{Key Management Protocol for the Sink Node}
\begin{figure}[H]
\centering
\includegraphics[width=11 cm]{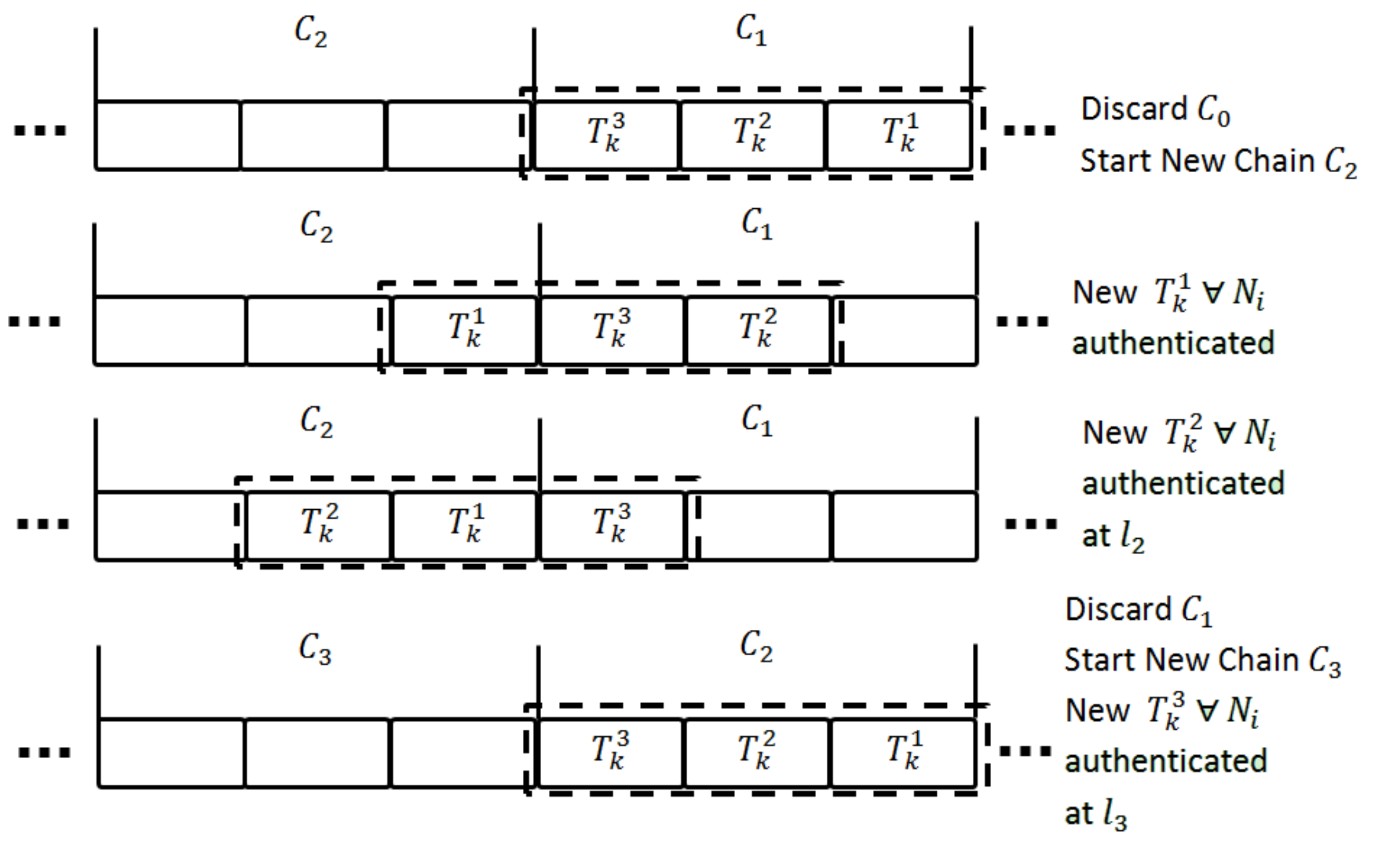}
\caption{Managing the chain of the key generator in the sink.}\label{fig4}
\end{figure}

At the start of the Chain of the Key Generator, the sink node reissues 
the ticket and session keys to associated nodes. To spread out the 
workload in the time dimension the sink node keeps the history of the 
previous key chain and issues a ticket/session key based on the moving 
window algorithm. Let us consider a chain with a length of 3; Figure \ref{fig4} 
shows how the sink node spread the workload throughout the chain by 
adopting the moving window approach. 

Step 1: The sink node discards the previous chain $C_{0}$ and 
generates the commitment key generator $G_{0}^{j} $ for the next chain 
$C_{2}$. Step 2: The sink node reissues the ticket and session key to 
all sensor nodes authenticated at interval 1. Step 3: The sink node 
reissues the ticket and session key to all sensor nodes authenticated at 
interval 2. Step 4: The sink node reissues the ticket and session key to 
all sensor nodes authenticated at interval 3 and discards the previous 
chain and generates the commitment key generator $ G_{0}^{j} $ for the 
next chain.

\subsection{Authentication Protocol Suite }
When a sensor node joins the system, it goes through the Sensor 
Activation and Authentication Protocol (SAAP). After SAAP, $N_{i}$ can 
establish multiple concurrent secure connections with sink nodes using 
the authentication ticket ($T_{k}$); similarly, $N_{i}$ uses the $
T_{k}$ for re-authentication while moving across the network. Likewise, 
when a user node $U_{i} $ joins the system, it goes through User 
Activation and Authentication Protocol (UAAP); subsequently, $U_{i}$ 
can use the authentication ticket ($T_{k}$) for authorization to 
collect data from multiple sink and sensor nodes. 
In a concurrent run of multiple instances of the protocol, the message 
authentication plays a critical part in preventing the replay attack and 
to achieve the objectives of the authentication protocol as defined in 
\cite{13,14}. In an SMSN message, authentication is accomplished by a 
secure exchange of a randomly generated nonce challenge. Moreover, in 
all the protocols discussed below, if the protocol initiator (user or 
sensor node) does not hear the response to an authentication/switch 
request, the protocol initiator resends the authentication/switch 
request including a new nonce and 'resend' flag. This step helps detect 
the impersonation, replay, and parallel session attack.

\subsubsection{Sensor Activation and Authentication Protocol (SAAP)}
Sink node $S_{j }$ periodically broadcasts a $Hello$ message ($
BS_{j}||S_{j} )$. If a node $N_{i}$ wants to join the WSN, upon 
hearing the $Hello message, N_{i}$ generates an encryption key $
K_{TS}^{i}=H(BS_{j}\oplus n_{S}^{i})$, encrypts the joining message 
with the generated key, and continues as follows: 
\begin{figure}[H]
\centering
\includegraphics[width=8 cm]{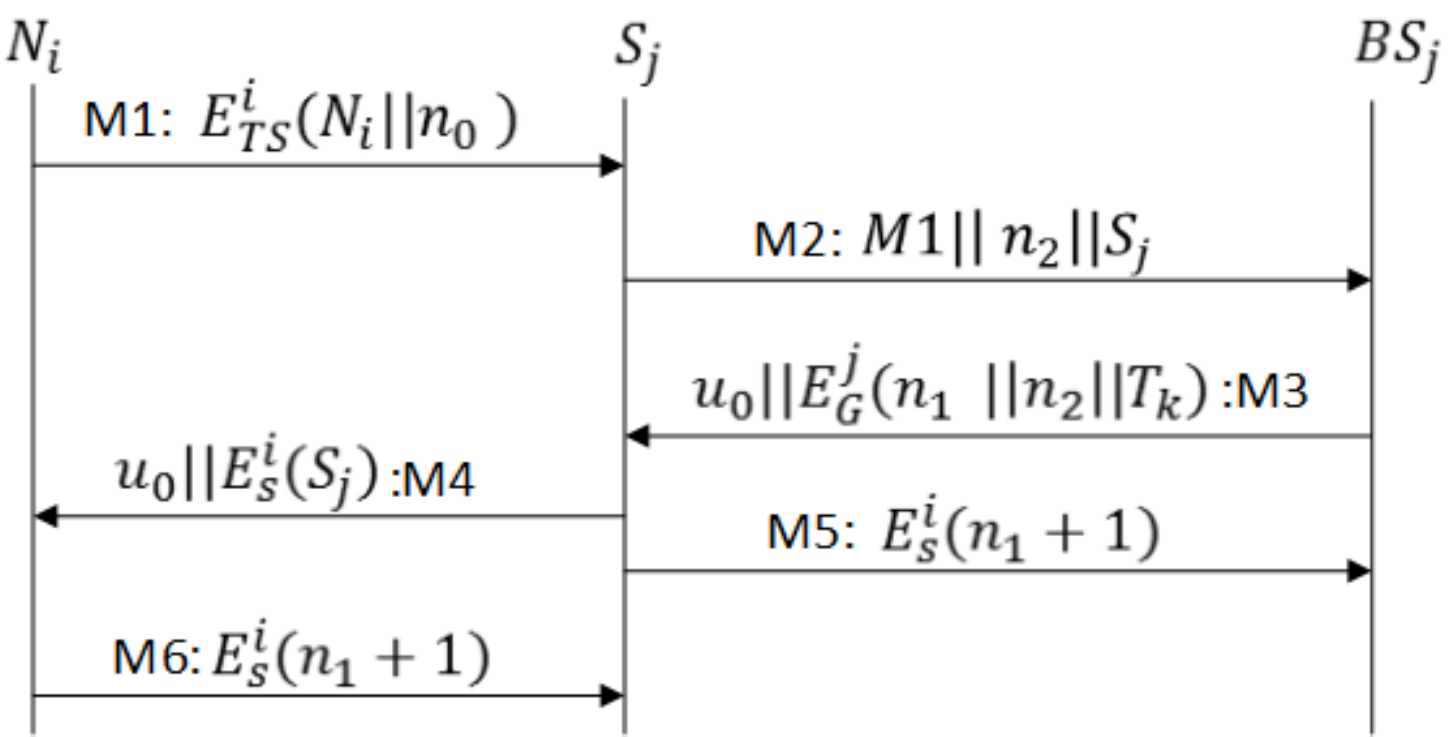}
\caption{Message exchange for Sensor Activation and Authentication Protocol (SAAP) .}\label{fig5}
\end{figure}

\begin{enumerate} [label=M\arabic*]
\item As shown in Figure \ref{fig5}, $N_{i}$ sends a JOIN message to sink $
S_{j}$ enclosing $ n_{0}$ and encrypted with the generated encryption 
key $K_{TS}^{i}$. 
\item Upon receiving the request from $N_{i}$, the $S_{j}$ forwards 
the request in conjunction with its identity and challenge $n_{2}$ to 
base station $BS_{j}$.
\item $BS_{j}$ retrieves the profile from the database, and if $
N_{i}$ isa legitimate sensor node, $BS_{j}$ generates the key $ 
K_{TS}^{i}=H(BS_{j}\oplus n_{S}^{i})$, and sends $u_{0}=E_{TS}^{i}$ $
(n_{0}+1||n_{1}||T_{k}||T_{R}||K_{s}^{i})$ to $S_{j}$ in M3. M3 
also includes ticket $T_{k}$, $n_{1}$ (a challenge for $N_{i}$
), and $n_{2}$ (challenge response for the sink node), all encrypted 
with $K_{G}^{j}$. The sink node $S_{j}$ verifies the challenge $ 
n_{2}$, stores $n_{1}$ and retrieves the profile and $K_{s}^{i}$ 
from the ticket. 
\item $S_{j}$\textit{ }forwards the \textit{ }$u_{0}$ to $
N_{i}.$\textit{ }After a challenge ($n_{0}+1)$ verification, $ N_{i} $ accepts $T_{k}$ and may start sending data to 
$S_{j}$.
\item $S_{j}$ sends the challenge response $(n_{1}+1)$ to $BS_{j}$ 
for the confirmation of a successful protocol run. 
\item After challenge ($n_{1}+1)$ confirmation, $S_{j}$ starts 
accepting sensor data; otherwise, it marks $T_{k}$ as an invalid 
ticket.

\end{enumerate}

In the SAAP, a secure exchange of $n_{0}$ ensures the message 
authentication between the sensor node and the base station, $n_{2}$ 
between the sink node and base station, and $n_{1} $ between the base 
station and the sink node, while message authentication between the 
sensor and sink nodes is established by session key encryption and $
n_{1}$. If $N_{i}$ is already registered with $BS_{j}$, in M1 $
n_{0}$ can be replaced with the hash of the password value. 

\subsubsection{Sensor Re-Authentication Protocol -1 (SRP1)}
If sensor node $N_{i}$ wants to establish multiple secure connections 
with sinks $S_{j}\in G_{i}$ or when a $N_{i}$ moves from $S_{k}\to 
S_{j}$ such that $ \lbrace S_{k},S_{j}\rbrace \in G_{i}$, the 
Re-Authentication Protocol -1 continues as follows: 
\begin{figure}[H]
\centering
\includegraphics[width=4.5 cm]{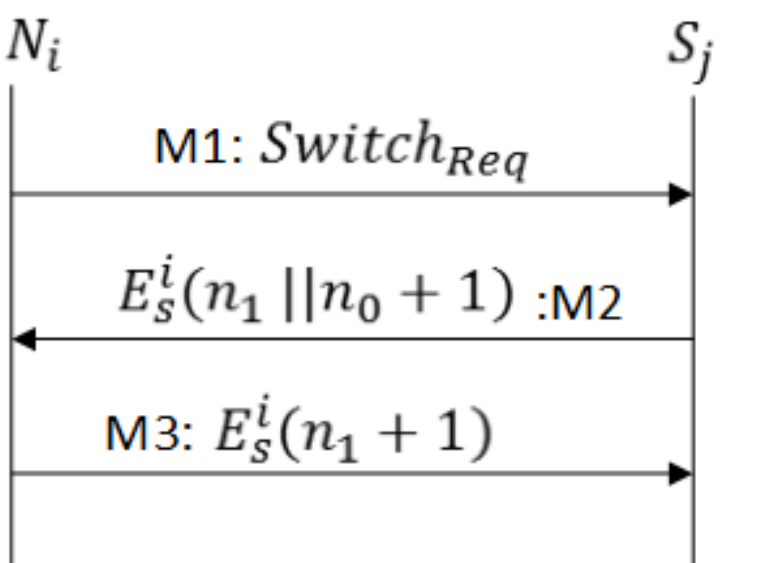}
\caption{Message exchange for Sensor Re-Authentication Protocol-1 (SRP1).}\label{fig6}
\end{figure}

\begin{enumerate}[label=M\arabic*]
\item As shown in Figure \ref{fig6}, $N_{i} $ sends $ 
Switch_{Req}=E_{S}^{i}(N_{i}||H(n_{0}^{i}))||T_{k}$ to $ S_{j}$. The 
sink $S_{j} $ decrypts the ticket, retrieves $K_{s}^{i},$ calculates 
the hash of $n_{0}^{i}$, and makes a comparison with $H(n_{0}^{i})$ 
received in the switch message. If the value $H(n_{0}^{i})$ does not 
match, the $S_{j}$ will ignore the request and otherwise proceed as 
follows.
\item $S_{j}$ sends a challenge response along with the new challenge 
encrypted with session key $K_{s}^{i}$. 
\item $N_{i} $ sends the challenge response $n_{1}$. After challenge 
confirmation, $S_{j}$ starts accepting data; otherwise, it marks $
T_{k}$ as an invalid ticket. 

\end{enumerate}

In the above procedure, secure exchange of $n_{0}^{i} $ ensures the 
message authentication between the sensor node and the sink node. With 
this feature, a sensor node can establish multiple secure sessions with 
various sink nodes as shown in Figure 1.

\subsubsection{Sensor Re-Authentication Protocol -2 (SRP2)}
If sensor node $N_{i}$ wants to establish another secure connection 
with sinks $S_{j}\in G_{k}^{o}$ or when a $N_{i}$ moves from $
S_{k}\to S_{j}$ such that $S_{j}\in G_{k}^{o}$, the re-authentication 
procedure proceeds as follows: 
\begin{figure}[H]
\centering
\includegraphics[width=8 cm]{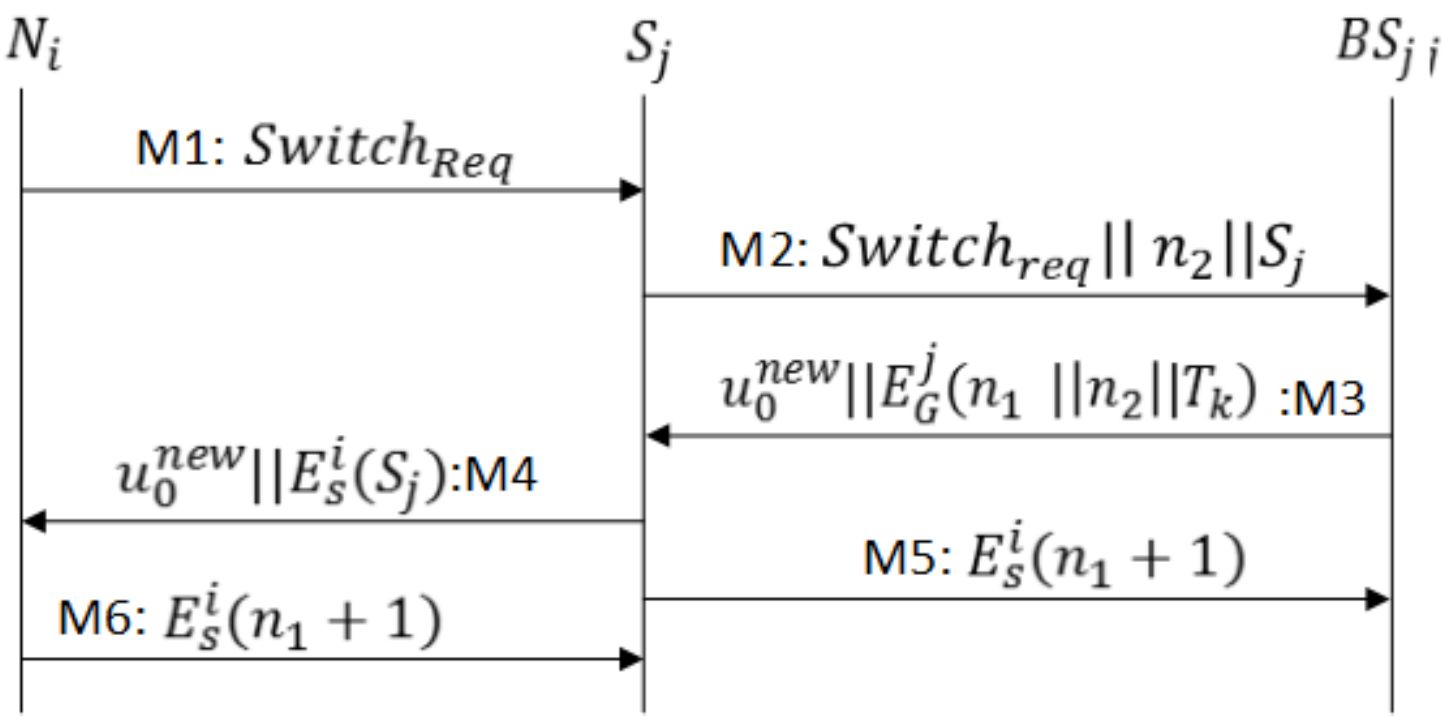}
\caption{Message exchange for Sensor Re-Authentication Protocol-2 (SRP2).}\label{fig7}
\end{figure}

\begin{enumerate}[label=M\arabic*]
\item As shown in Figure \ref{fig7}, $N_{i} $ sends $
Switch_{Req}=E_{S}^{i}(N_{i}||H(n_{0}^{i}))||T_{k}$ to $ S_{j}$. The 
sink $S_{j} $ decrypts the second half of $T_{k}$ and verifies the 
identities of $N_{i}$ and ticket granting base station. 
\item After identities verification the sink $S_{j}$ forwards the 
request in conjunction with a challenge $n_{2}$ to base station $
BS_{j}$. 
\item $BS_{j}$ retrieves the profile from $T_{k}$, and if $N_{i}
$ isa legitimate sensor node, the $BS_{j}$ generates the key $ 
K_{TS}^{i}=H(BS_{j}\oplus n_{S}^{i})$, and sends $
u_{0}^{new}=E_{TS}^{i}(H(n_{0}^{i})||n_{1}||T_{k}||T_{R}||K_{s}^{i})
$ to $S_{j} $ in M3. M3 also includes ticket $T_{k}$, $n_{1}$ (a 
challenge for $N_{i}$), and $n_{2}$ (challenge response for the 
sink node), all encrypted with $K_{G}^{j}$. The sink node $S_{j}$ 
verifies the challenge $ n_{2}$, stores $n_{1}$ and retrieves the 
profile and $K_{s}^{i}$ from the ticket. 
\item $S_{j}$\textit{ }forwards\textit{ }$u_{0}^{new}$ to $
N_{i}.$\textit{ }After challenge ($n_{0}+1)$ verification $ N_{i} 
$ accepts $T_{k}$ and start sending data to $S_{j}$.
\item $S_{j}$ sends the challenge response $(n_{1}+1)$ to $BS_{j}$
; it confirms a successful protocol run.
\item After challenge ($n_{1}+1)$ confirmation $S_{j}$ start 
accepting data; otherwise, it marks $T_{k}$ as an invalid~ticket. 
\end{enumerate}

In the SRP2, a secure exchange of $n_{0}$ ensures the message 
authentication between the sensor node and the base station, $n_{2}$ 
between the sink node and base station, and $n_{1} $ between the base 
station and the sink node, while message authentication between the 
sensor and sink nodes is established by session key encryption and $
n_{1}$. If $N_{i}$ wants to share data in a secret mode, both $
N_{j}$ and $S_{j}$ generate a private session key $
K_{sp}^{j}=H(K_{s}^{i},n_{0}^{i},n_{1})$. With this feature, a sensor 
node can establish multiple secure sessions with various sink nodes as 
shown in Figure \ref{fig1}.

\subsubsection{User Activation and Authentication Protocol (UAAP)}
In some scenarios a user may desire to access data from the sensor 
network, for example, in IoT applications such as smart homes and smart 
buildings, a smartphone user may want to get sensor node data. In SMSN 
authentication protocol suites, an authenticated user (holding a valid 
ticket) can access data directly from sensor nodes and/or can collect 
from the sink node. The ticket structure for $U_{i}$ is the same as 
the ticket structure discussed above for $N_{i}$ where the sensor node 
identity and profile are replaced with the user node identity and 
profile. The user profile information includes the permissible 
accessibility information. 
In the User Activation and Authentication Protocol (UAAP) a user can 
acquire a ticket from $BS_{j}$ in two different ways. If the user is 
not in the communication range of $BS_{j}$ it routes the joining 
message via the nearest sink node $S_{j }$; the protocol proceeds 
exactly as in the SAAP. However, if the user is in the communication 
range of $BS_{j}$, the user $ U_{i}$ generates an encryption key $
K_{TS}^{i}=H(BS_{j}\oplus n_{S}^{i})$, encrypts the joining message 
with the generated key, and proceeds as follows: 
\begin{figure}[H]
\centering
\includegraphics[width=4.5 cm]{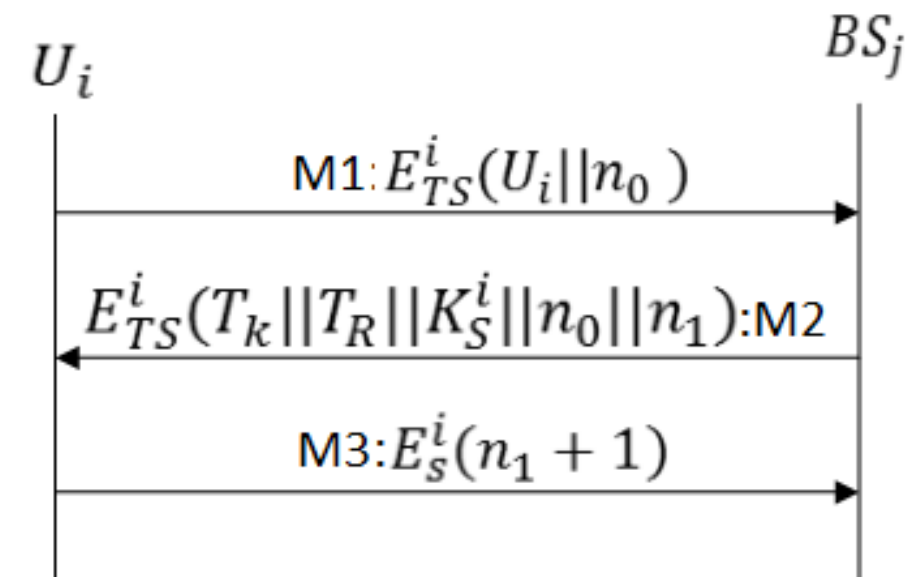}
\caption{Message exchange for User Activation and Authentication Protocol (UAAP).}\label{fig8}
\end{figure}

\begin{enumerate}[label=M\arabic*]
\item As shown in Figure \ref{fig8}, $U_{i}$ sends a JOIN message enclosing$ 
n_{0}$ and the user identity to base station~$BS_{j}$.
\item $BS_{j}$ retrieves profile from the database, and if $U_{i}$
is a legitimate user, the $BS_{j}$ generates authentication ticket, 
and send along with, $n_{1}$ ( a challenge for $U_{i}$), and $
n_{0}$ (challenge response), all encrypted with $ 
K_{TS}^{i}=H(BS_{j}\oplus n_{S}^{i})$. The ticket structure is same 
except the secret nonce $n_{1}$ enclosed inside ticket is generated by~$BS_{j}. $
\item The user node $U_{i}$ verifies the challenge $ n_{0}$, stores 
$T_{k}$ and $ n_{1}$. $U_{i}$ sends the challenge response $
(n_{1}+~1)$ to $ BS_{j}$; it confirms a successful protocol run . 
After receiving a challenge response $(n_{1}+1)$ the $BS_{j}$ 
updates the status of $U_{i} $ from idle to active user.

\end{enumerate}

In User Activation and Authentication Protocol, secure exchange of $
n_{0}$ ensures the message authentication between user node and base 
station and $n_{1} $between the base station and user node.

\subsubsection{User-Sink Authentication Protocol (USiAP) }
After acquiring the authentication ticket, if user $U_{k} $ wants to 
retrieve data from sink nodes $S_{j}$, it sends a JOIN request in 
conjunction with a ticket to $S_{j}$ and then follows the same 
procedure as discussed in Sensor Re-Authentication Protocols 1 and 2; 
except after ticket verification, $S_{j}$ can piggyback data with the 
rest of the messages. Using the authentication ticket, user $U_{k}$
can also establish multiple concurrent connections with various sink 
nodes. 

\subsubsection{User- Sensor Authentication Protocol (USeAP)}
After acquiring the authentication ticket, if user $U_{k} $ wants to 
access data directly from sensor nodes $ N_{i}$, it sends a JOIN 
request in conjunction with a ticket to $N_{i}$ and the procedure 
proceeds as~follows:
\begin{figure}[H]
\centering
\includegraphics[width=8 cm]{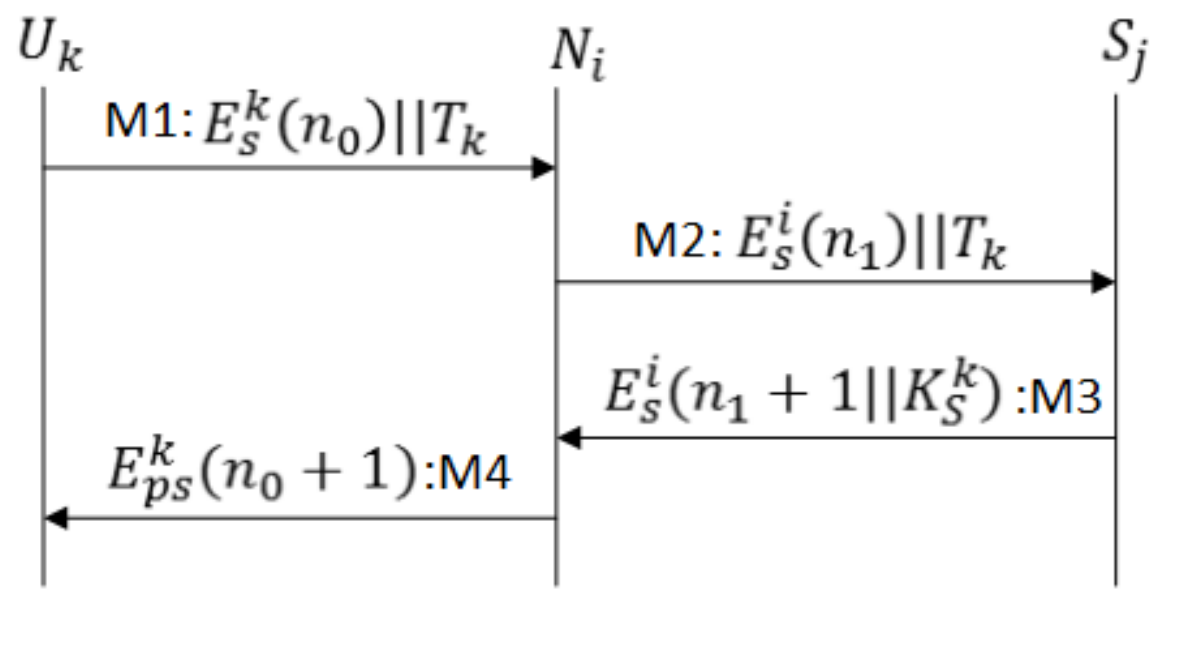}
\caption{Message exchange for User-Sensor Authentication Protocol (USeAP).}\label{fig9}
\end{figure}

\begin{enumerate}[label=M\arabic*]
\item As shown in Figure \ref{fig9}, $U_{k}$ sends a JOIN message in 
conjunction with ticket and challenge $n_{0}$ encrypted with its 
ticket centered session key.
\item Upon receiving the request from $U_{k}$ the $N_{i}$ forward 
the ticket and encrypted challenge $n_{1}$ to sink node $s_{j}.$
\item Sink decrypts the ticket and retrieves $U_{k}'$s profile and 
session key $K_{S}^{k}$, and if $U_{i}$ is a legitimate user node, 
$S_{j}$ sends $K_{S}^{k}$ and challenge response $(n_{1}+1)$ to $
N_{i}$.
\item After challenge verification, $ N_{ i}$ generates a private 
session key $K_{ps}^{k}=H(K_{s}^{k},n_{0})$ and sends a challenge 
response encrypted with the private session key. $U_{k}$ also 
generates the private session key and verifies the challenge response.
\end{enumerate}

In User-Sensor Authentication Protocol suite, secure exchange of $n_{0}$ 
ensures the message authentication between users and sensor node, while 
the secure exchange of $n_{1}$ ensures the message authentication 
between sensor node and sink node.

\section{Security Analysis}
This section presents the comprehensive security analysis of the SMSN 
protocol, including an informal security analysis and discussion of a 
formal security analysis using BAN logic \cite{32}, and finally presents 
the Scyther \cite{11,12} implementation result of the SMSN and previously 
proposed schemes~\cite{16,17,18,19,20,21}.

\subsection{Informal Analysis and Discussion}
To verify the strength of the SMSN protocol against known attacks we 
introduce an intruder in the network with capabilities as follows: It 
has an initial information set that contains the IDs of all users, 
sensor nodes, sink nodes and base stations. It can intercept and record 
message exchanges between participating entities. It can redirect, 
spoof, and replay the messages. The subsequent sections show that the 
intruders, with all the above-mentioned capabilities, fail to launch a 
successful replay, parallel session, man-in-middle , impersonation, and 
several other attacks against the SMSN protocol suite. 

\subsubsection{Replay, Multiplicity, Parallel and Man in Middle Attacks 
Against the SMSN}
We introduce an intruder, as discussed above, in the network and launch 
replay, multiplicity, parallel session, and man-in-middle attacks 
against the SMSN for three different scenarios, as shown in Figures 10, 
11 and 12. The intruder $Z $impersonates a protocol participant, 
intercepts the messages, and replays them to deceive other protocol 
participants. Replay, multiplicity, parallel, and man-in-middle attacks 
against the SAAP for three different scenarios are given below.

\textbf{Scenario 1:} 

\begin{figure}[H]
\centering
\includegraphics[width=9 cm]{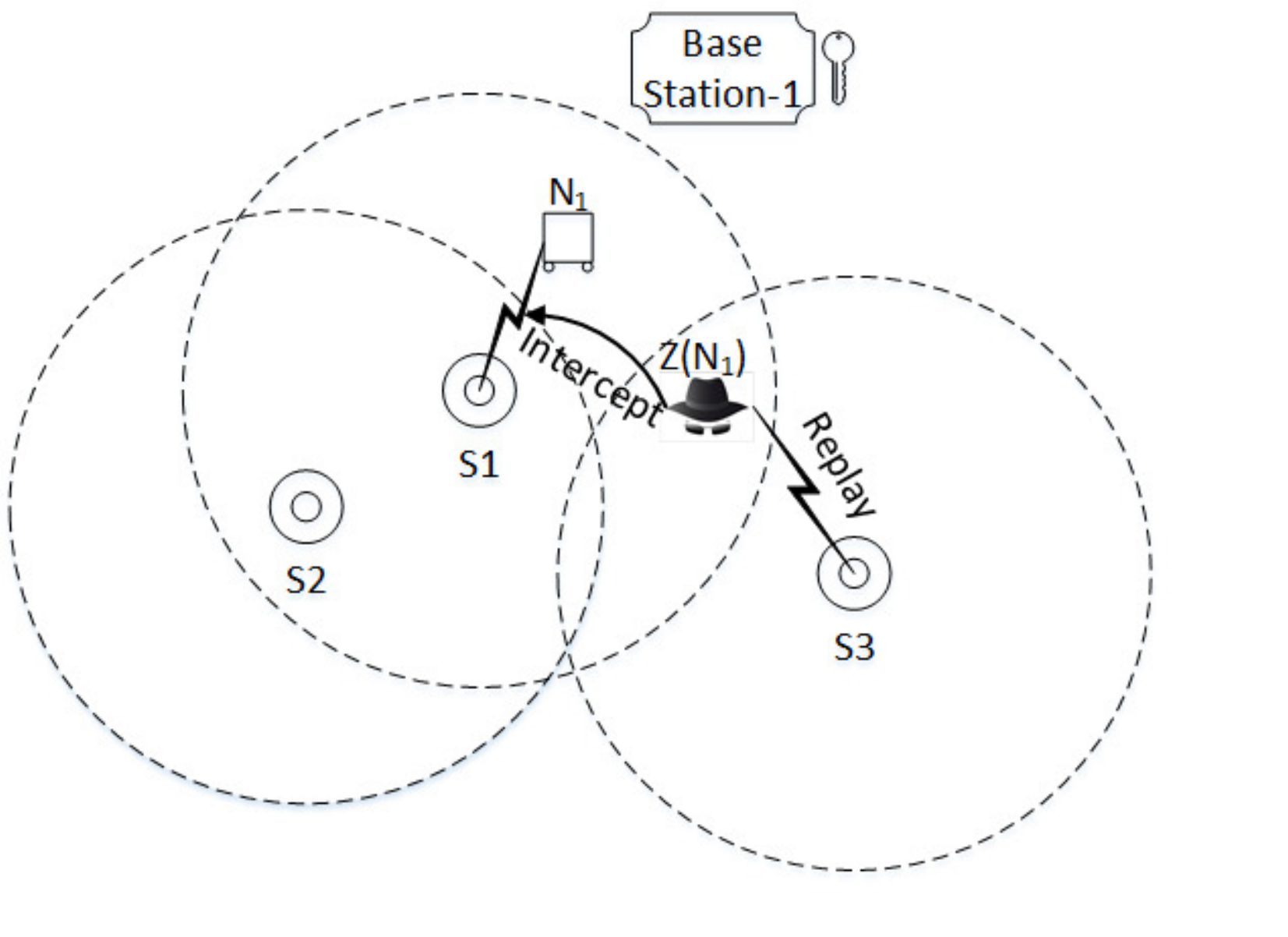}
\caption{The intruder $Z$ impersonates the protocol initiator $N_1$ (can be a sensor or user node). Intruder $Z(N_1)$ intercepts the messages between $N_1$ and sink $S1$ and replays them to sink $S2$. Both sinks are associated with base station 1 and share the same keychain.}\label{fig10}
\end{figure}

For the given scenario in Figure \ref{fig10}, let us suppose a sensor node $ 
N_{i}$ sends a request for authentication to sink $S_{j}\sim BS_{j}$
. During the protocol run an intruder $Z(N_{i})$ intercepts the 
messages and replays them to another sink $S_{i}\sim BS_{j}$; the 
attack proceeds as follows: The intruder $Z(N_{i}) $ intercepts M1 and 
replays it to $S_{i}$. The attack is detected immediately when $
BS_{j}$ receives two M2 messages enclosing the same M1. $BS_{j}$ 
sends M3 to both sinks comprising the 'Alert' flag and a different $ 
n_{1}$ nonce challenge. $Z(N_{i})$ intercepts M6 and replays it to $
S_{i}$; note that the intercepted message M6 comprises a different $
n_{1}$ which is the only valid response for a sink $S_{j}$. Upon 
receiving the wrong challenge response, the sink $S_{i}$ identifies 
the intruder node. The SAAP not only detects the replay attack but also 
identifies the intruder. 

\textbf{Scenario 2:} 

\begin{figure}[H]
\centering
\includegraphics[width=9 cm]{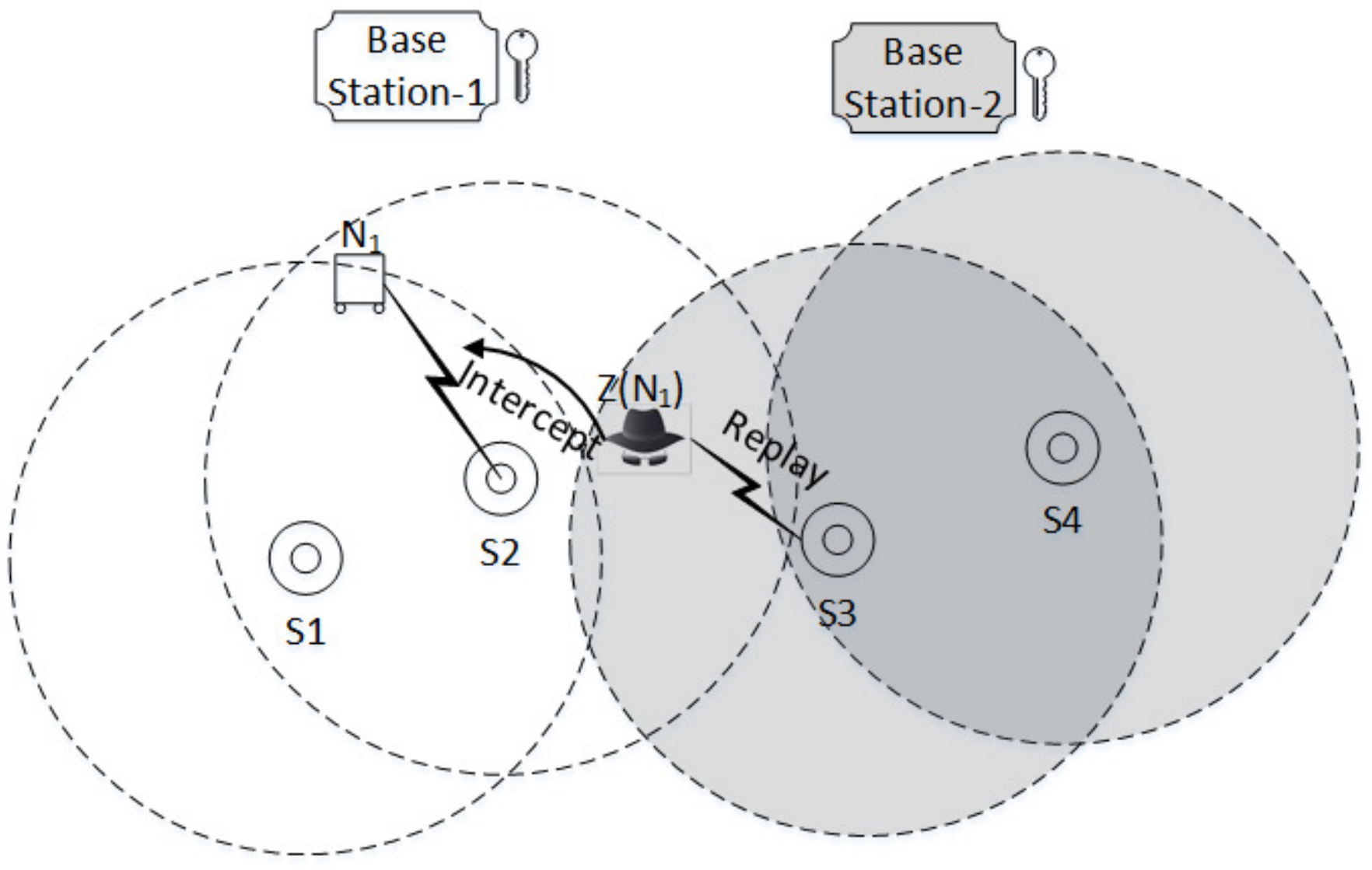}
\caption{The intruder $Z$ impersonates the protocol initiator $N_1$ (can be a sensor or user node). Intruder $Z(N_1)$ intercepts the messages between $N_1$ and sink $S2$ and replays them to sink $S3$. Sink $S2$ is associated with base station 1 , and Sink $S3$ is associated with base station 2. }\label{fig11}
\end{figure}

For the given scenario in Figure \ref{fig11}, let us suppose a sensor node $ 
N_{i}$ sends a request for authentication to sink $S_{j}\sim BS_{j}$
. During the protocol run an intruder $Z(N_{i})$ intercepts the 
messages and replays them to another sink $S_{i}\sim BS_{i}$; the 
attack proceeds as follows: The intruder $Z(N_{i}) $ intercepts M1 and 
replays it to $S_{i}$. Unlike in scenario 1, neither base station $
BS_{j}$ and $BS_{i}$ can detect the attack at this stage and replies 
with a normal M3 to associated sink nodes $S_{i}$ and $S_{j}, $
respectively. However, both M3 messages comprise a different $n_{1}$ 
nonce challenge. $Z(N_{i})$ intercepts the M6 and replays it to $
S_{i}$; note that the intercepted message M6 contains a different $
n_{1}$ which is the only valid response for the sink $S_{j}$. Upon 
receiving the wrong challenge response, the sink $S_{i}$ identifies 
the intruder. 

\textbf{Scenario 3:} 

\begin{figure}[H]
\centering
\includegraphics[width=9 cm]{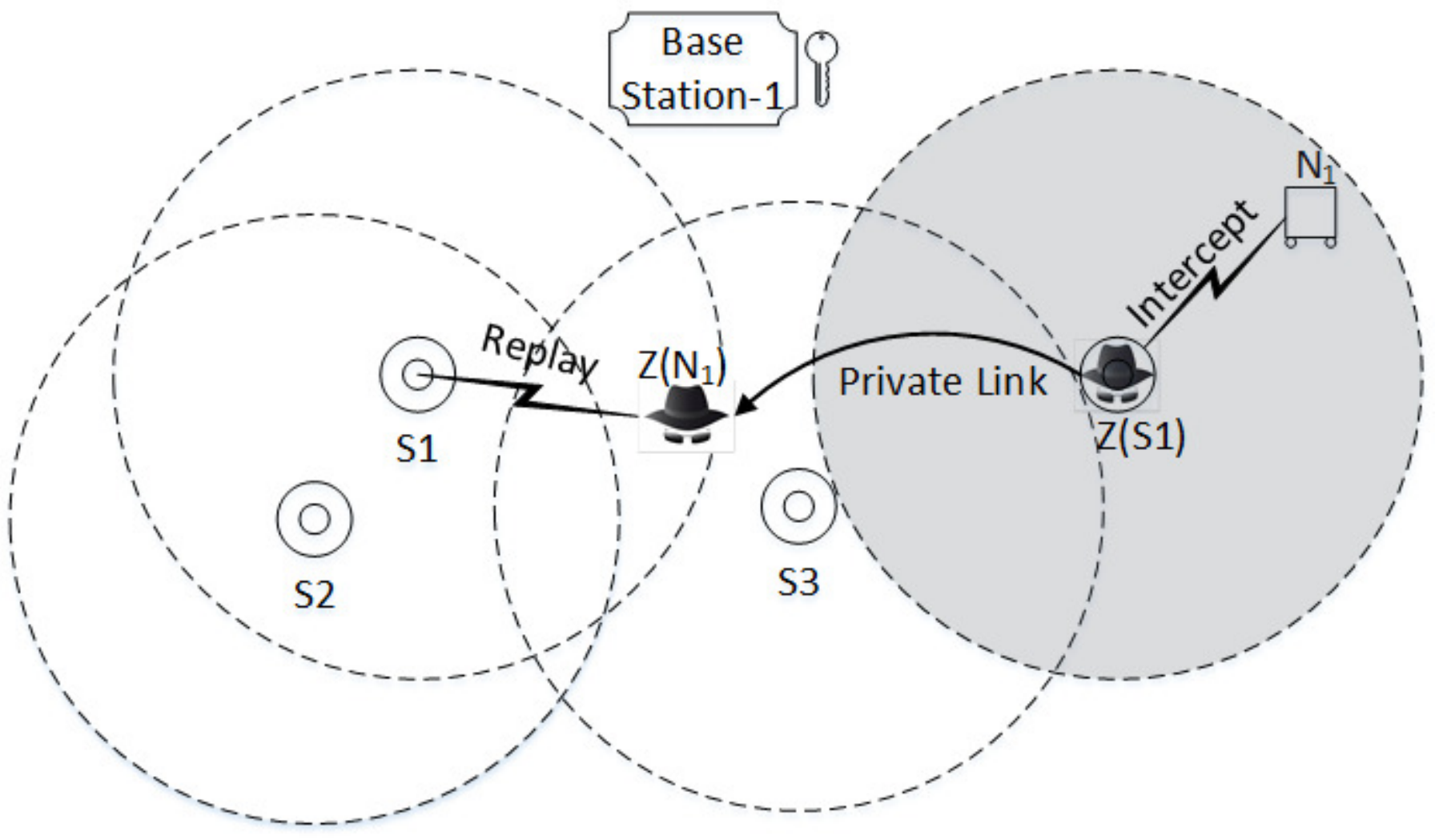}
\caption{Two intruders impersonate the sink node $S_1~BS_1$ and sensor node $N_1$; the intruder $Z(N_1)$ is within the region of $S_1~BS_1$, and intruder $Z(S_1)$ is outside somewhere close to node $N_1$. Both intruders can communicate with zero communication delay via a private link. }\label{fig12}
\end{figure}

For the given scenario in Figure \ref{fig12}, let us suppose two intruders 
impersonate the sink $S_{j}\sim BS_{j}$ and sensor node $N_{i}$; 
the intruder $Z(N_{i})$ is within the region of $S_{i}\sim BS_{i}$ , 
and intruder $Z(S_{j})$ is outside somewhere close to node $N_{i}.$ 
Furthermore, both intruders can communicate through a private link with 
zero delay. During the protocol run intruder $Z(S_{j})$ intercepts the 
messages sent by sensor node $N_{i}$ and replays it to $BS_{j}$; 
also $Z(S_{j})$ shares all the intercepted messages with fellow 
intruder $Z(N_{i})$ via a private secure channel; the attack proceeds 
as follows: The intruder $Z(S_{j})$ intercepts M1 sent by $N_{i}$ 
and shares the intercepted message with $Z(N_{i})$. The intruder $
Z(N_{i})$ sends the intercepted message to $S_{i}\sim BS_{i}$; the 
protocol proceeds normally and upon receiving M4 the intruder $Z(N_{i})
$ sends the message M4 to $Z(S_{j})$. The intruder $Z(S_{j})$ sends 
M4 to sensor node $N_{i}$; upon receiving M6 the intruder $Z(S_{j})
$ shares the message M6 with $Z(N_{i})$. The intruder $Z(N_{i})$ 
sends M6 to $S_{i}\sim BS_{i}$ and the attack is completed. The attack 
is only successful if the replay of M6 is delivered to $S_{j}$ within 
a time interval of $Td/L$ where $L$ is the length of the keychain. 
Moreover, in practice when the private link between fellow intruders 
adds a communication delay, $N_{i}$ can detect the attack by comparing 
the $TR$ (registration time sent in $u_{0}) $ with the local time.

As the outcome of the above attack, $N_{i}$ considers that the 
authentication process was completed successfully. However, the 
intruders cannot get any useful information during a protocol run: $
Z(S_{j})$ does not know the session key delivered to $N_{i}$, so $ 
Z(S_{j})$ cannot further communicate with $N_{i}$. Due~to the 
unavailability of a data link, $N_{i}$ uses the ticket to run SRP1/2. 
However, a problem exists on the other side: sink $S_{i}$ and base 
station $BS_{i}$ consider that they authenticated a legitimate sensor 
$N_{i}$ successfully; from their point of view, the sensor $N_{i}$ 
is within the region of $BS_{i}$ but in reality, $N_{i}$ is in the 
region of $BS_{j}$. Similarly, in the case of SRP1, SRP2, UAAP, 
USiAP, and USeAP the intruder $Z$ fails to launch successful attacks 
for scenarios 1 and 2; however, the intruder $Z $completes the attack 
in scenario 3.

In a nutshell, replay, multiplicity, parallel and man-in-middle attacks 
against the SMSN protocol are not successful. Even though in scenario 3 
the intruders completed the attack, they could not cause a serious 
security issue because after the completion of the protocol run, the 
intruder could not get useful information such as the session key, 
ticket, or partial session key. Furthermore, the attack can be avoided 
by introducing a timestamp in each message exchange. For illustration, 
the intruders need to wait for a significantly longer time to replay M4 
and M6, and this significant delay can be detected with the time stamp, 
which reveals the existence of the intruder in the network. 

\subsubsection{Black Hole Attack}
In a black hole attack \cite{46,47,48} the intruder impersonates a node and 
blocks or drops the messages upon receiving them. In the SMSN, the 
sensor and user nodes can connect to multiple sink nodes simultaneously; 
hence, failure of data exchange on one route does not block the data 
delivery towards the base station. Moreover, the black hole attack is 
detectable in our scheme because the SMSN ensures binding by employing 
an exchange of secret nonce between $ N_{i}\leftrightarrow S_{j}, 
N_{i}\leftrightarrow BS_{j}$, and $B_{j}\leftrightarrow B.$ 
Consecutive failures of exchange of challenge detects the black hole 
attack. Once the black hole is detected, the sensor node can send data 
via another sink node. 

\subsubsection{Wormhole Attack}\textbf{\textit{ }}
\textit{ }In a wormhole attack \cite{49,50}, the intruder captures the 
messages in one location and tunnel to another location to a fellow 
intruder who replays the tunneled messages in another location area. The 
attack discussed in scenario 3 can be regarded as a\textbf{ }wormhole 
attack. From the point of view of replay, multiplicity, parallel and 
man-in-middle attacks, the attack in scenario 3 did not achieve its 
objectives, but from the point of view of the wormhole attack, the 
attack is successful. The solution for the problem is similar to the one 
we discussed earlier: the attack can be avoided by introducing a 
timestamp in each message exchange. 

\subsubsection{Analytical Attacks}
In an analytic attack \cite{31,52}, the intruder intercepts the messages 
and using cryptanalysis tries to recover a cryptographic key. With the 
inclusion of a time-based key, our scheme inherits the freshness 
property, which defies the capability of the intruder to launch 
analytical attacks, as it has a max time of $T_{d}$ to acquire the 
time-based key, which makes it difficult to launch analytical attacks. 

\subsubsection{Topological Centered Attacks}
In \cite{53,54} the authors presented an authentication protocol for a 
sensor network in which the sink issues a re-authentication ticket that 
includes a list of neighbor sink nodes. This information can lead to 
topological centered attacks \cite{27,28} such as identity replication 
attacks. In our scheme, the topological information is entirely obscured 
from the sink and sensor nodes. 

\subsection{Formal Analysis Using BAN Logic}

The BAN logic \cite{32} is a widely used formal method for the formal 
analysis of security protocols. To prove the security of the SMSN 
protocol suite it is sufficient to demonstrate the security of the SAAP 
and UAAP protocols; the rest of the protocols are extensions of the SAAP 
and UAAP and use the ticket and session key established in the SAAP and 
UAAP protocols run. Hence, proof of SAAP and UAAP protocols concludes 
the security of the SMSN protocol suite. 

The three basic objects of BAN logic are principals, formula/statements, 
and encryption keys. The~principals, the protocol participants, are 
represented by symbols $P$ and $Q$. The formula/statements are 
symbolized by $X$ and $Y$ and represents the content of the message 
exchanged. The encryption keys are symbolized by $K$. The logical 
notations of BAN-logic used for our analysis is given below:

\begin{itemize}[leftmargin=2.7em,labelsep=5mm]
\item $P\models X:P$ believes $X$, or $P $ would be enabled to believe $
X$; in conclusion, $P$ can take $X $ as true.
\item $P \prec X:P $ sees/receives $X$. $P $ initially has or received a 
message $X $ and $P$ can see the contents of the message and is 
capable of repeating $X$.
\item $P|\sim X:P$ once said $X.$ $P $ has sent a message including 
the statement $X$. However, the freshness of message is unknown.
\item $P\Rightarrow X:P $ controls $X$ and should be trusted for 
formula/statement $X$.
\item $\#(X):X$ is fresh; it says, $X $ never sent by any principal 
before. 
\item $P\xleftrightarrow{ K }Q:P $ and $Q$ shares a key $K$ to 
communicate in a secure way and $K$ is only known to $P$, $Q$ and 
a trusted principal.
\item $(X)_{K}: $ The statement $X$ is encrypted by key $K.$
\item $\lbrace X\rbrace _{Y}: $ It stand for $X $ combined with $Y.$ 
$Y $ is anticipated to be secret and its implicit or explicit presence 
proves the identity of a principal who completes the $\lbrace X\rbrace 
_{Y}$.
\end{itemize}

Some primary BAN-logic postulates used in the analysis of the SMSN are 
given below:

\begin{itemize}[leftmargin=2.7em,labelsep=5mm]
\item Message meaning rules: $\frac{P\models P\xleftrightarrow{ K 
}Q,P\prec (X)_{K} }{P\models Q|\sim X}$ ,$\frac{P\models P\xleftrightarrow{ Y 
}Q,P\prec {X}_{Y} }{P\models Q|\sim X} $ 
\item Nonce verification rule: $\frac{P\models \#(X),P\models Q|\sim X }{P\models Q\models X}$ 
\item Jurisdiction rule: $\frac{P\models Q\Rightarrow X,P\models Q\models X }{P\models X}$ 
\item Freshness rule: $\frac{P\models \#(X)}{P\models (X,Y)}$ 
\item Believe rule: $\frac{P\models Q\models (X,Y) }{P\models X,P\models Y}$ 
\item Session key rule: $\frac{P\models Q\#(X),P \models Q\models X }{P\models P\xleftrightarrow{ 
K }Q}$ 
\end{itemize}

\subsubsection{BAN Logic Analysis of SAAP}
The SAAP protocol should achieve the following goals:

\begin{enumerate}[label=G\arabic*]
\item $N_{i}\models (N_{i}\xleftrightarrow{ K_{S}^{i} }S_{j})$
\item $N_{i}\models S_{j}\models (N_{i}\xleftrightarrow{ K_{S}^{i} }S_{j})$
\item $S_{j}\models (N_{i}\xleftrightarrow{ K_{S}^{i} }S_{j})$
\item $S_{j}\models N_{i}\models (N_{i}(N_{i}\xleftrightarrow{K_{S}^{i}}S_{j})$
\end{enumerate}

Protocol Idealization:

\begin{enumerate}[label=I\arabic*]
\item $N_{i}\xleftrightarrow{Via S_{j}}BS_{j}:\lbrace 
n_{0},(N_{i}\xleftrightarrow{ n_{S}^{i} }BS_{j})\rbrace _{H(X_{S})}
$ 
\item $S_{j}\to BS_{j}:\lbrace n_{2}, IDS_{j}\rbrace $
\item $BS_{j}\to S_{j}:\lbrace (n_{1}, n_{2}, (N_{i}\xleftrightarrow{ K_{S}^{i} }S_{j}, \# (N_{i}\xleftrightarrow{ K_{S}^{i} 
}S_{j}),n_{0}^{i})_{K_{l}^{i}=H(\zeta 
_{l}^{j},H(n_{0}^{i}) )},V_{i})_{K_{G}^{i}},\lbrace 
n_{0},n_{1},(N_{i}\xleftrightarrow{ K_{S}^{i} }S_{j}, 
\# (N_{i}\xleftrightarrow{ K_{S}^{i} 
}S_{j}),n_{0}^{i})_{K_{l}^{i}=H(\zeta 
_{l}^{j},H(n_{0}^{i}))},T_{R},N_{i}\xleftrightarrow{ K_{S}^{i} 
}S_{j},\# (N_{i}\xleftrightarrow{ K_{S}^{i} 
}S_{j}),N_{i}\xleftrightarrow{ n_{S}^{i} }BS_{j}\rbrace 
_{H(X_{S})}\rbrace $ 
\item $S_{j}\xleftrightarrow{Via BS_{j}} N_{i}:\lbrace 
n_{0},n_{1},(N_{i}\xleftrightarrow{ K_{S}^{i} }S_{j}, 
\# (N_{i}\xleftrightarrow{ K_{S}^{i} 
}S_{j}),n_{0}^{i})_{K_{l}^{i}=H(\zeta 
_{l}^{j},n_{0}^{i})},T_{R},N_{i}\xleftrightarrow{ K_{S}^{i} 
}S_{j},\# (N_{i}\xleftrightarrow{ K_{S}^{i} 
}S_{j}),N_{i}\xleftrightarrow{ n_{S}^{i} }BS_{j}\rbrace _{H(X_{S})}
$ 
\item $S_{j}\to N_{i}:(IDS_{j})_{K_{s}^{i}}$
\item $N_{i}\to S_{j}:(n_{1})_{K_{S}^{i}}$

\end{enumerate}

Initial State Assumptions:
\begin{enumerate}[label=A\arabic*]
\item $BS_{j}\models \# (n_{0})$
\item $BS_{j}\models \# (n_{2})$
\item $S_{j}\models \# (n_{1})$
\item $N_{i}\models \# (n_{1})$
\item $N_{i}\models (N_{i}\xleftrightarrow{K_{TS}=H(X_{S}) }BS_{j})$ 
\item $BS_{j}\models (N_{i}\xleftrightarrow{K_{TS}=H(X_{S}) }BS_{j})$ 
\item $S_{j}\models (S_{j}\xleftrightarrow{K_{G}^{j} }BS_{j})$ 
\item $BS_{j}\models (S_{j}\xleftrightarrow{K_{G}^{j} }BS_{j})$
\item $BS_{j}\models N_{i}\models (N_{i}\xleftrightarrow{K_{TS}=H(X_{S}) }BS_{j})
$
\item $N_{i}\models BS_{j}\models (N_{i}\xleftrightarrow{K_{TS}=H(X_{S}) }BS_{j})
$
\item $BS_{j}\models S_{j}\models (S_{j}\xleftrightarrow{K_{G}^{j} }BS_{j})$
\item $S_{j}\models BS_{j}\models (S_{j}\xleftrightarrow{K_{G}^{j} }BS_{j})$
\end{enumerate}

Let us analyze the protocol to show that $N_{i}$ and $S_{j}$ share a 
session key:

From I1, we have
\begin{equation}
{BS_{j}\prec \lbrace n_{0},(N_{i}\xleftrightarrow{ n_{S}^{i} 
}BS_{j})\rbrace _{H(X_{S})}}\tag{1}
\end{equation}

The (1), A6 and message meaning rule infers that
\begin{equation}
{S_{j}\models N_{i}|\sim \lbrace n_{0},(N_{i}\xleftrightarrow{ n_{S}^{i} 
}BS_{j})\rbrace}\tag{2}
\end{equation}

The A1 and freshness conjuncatenation comprehends that
\begin{equation}
BS_{j}\models \# \lbrace n_{0},(N_{i}\xleftrightarrow{ n_{S}^{i} 
}BS_{j})\rbrace \tag{3}
\end{equation}

The (2), (3) and nonce verification rule deduces that
\begin{equation}
BS_{j}\models \lbrace N_{i}\models n_{S}^{i}, n_{0},(N_{i}\xleftrightarrow{ 
n_{S}^{i} }BS_{j})\rbrace \tag{4}
\end{equation}

The (4) and believe rule infers that
\begin{equation}
BS_{j}\models N_{i}\models (N_{i}\xleftrightarrow{ n_{S}^{i} }BS_{j}) \tag{5}
\end{equation}

From A2, (5) and jurisdiction rule, it concludes
\begin{equation}
BS_{j}\models (N_{i}\xleftrightarrow{ n_{S}^{i} }BS_{j}) \tag{6}
\end{equation}

This belief confirms that $BS_{j}$ has received a message from a 
legitimate$N_{i}$.

From I2, we have
\begin{equation}
BS_{j}\prec n_{2} \tag{7}
\end{equation}

The (7) and message meaning it infers that
\begin{equation}
BS_{j}\models S_{j}|\sim n_{2} \tag{8}
\end{equation}

The A2, A1, (3) and freshness conjuncatenation comprehends that

\begin{equation}
BS_{j}\models \# \lbrace n_{0},n_{2},(N_{i}\xleftrightarrow{ n_{S}^{i} 
}BS_{j})\rbrace \tag{9}
\end{equation}

According to nonce freshness, this proves that $BS_{j}$ confirms that 
$N_{i}$ is recently alive and running the protocol with $BS_{j}$.

From I3, we have
\begin{equation}
S_{j}\prec (n_{1}, n_{2},(N_{i}\xleftrightarrow{ K_{S}^{i} 
}S_{j},\# (N_{i}\xleftrightarrow{ K_{S}^{i} 
}S_{j}),n_{0}^{i})_{K_{l}^{i}=H(\zeta 
_{l}^{j},H(n_{0}^{i}))},V_{i})_{K_{G}^{i}} \tag{10}
\end{equation}

The A7 and (10) deduce that
\begin{equation}
S_{j}\models BS_{j}|\sim \lbrace n_{1}, n_{0}^{i},N_{i}\xleftrightarrow{ 
K_{S}^{i} }S_{j},\# (N_{i}\xleftrightarrow{ K_{S}^{i} 
}S_{j}),V_{i}\rbrace \tag{11}
\end{equation}

The A3, (11) and freshness conjuncatenation comprehends that

\begin{equation}
S_{j}\models \# \lbrace n_{1}, n_{0}^{i},N_{i}\xleftrightarrow{ K_{S}^{i} 
}S_{j},\# (N_{i}\xleftrightarrow{ K_{S}^{i} }S_{j}),V_{i}\rbrace \tag{12}
\end{equation}

The (11), (12) and nonce verification rule infers that
\begin{equation}
S_{j}\models BS_{j}\models \lbrace n_{1}, n_{0}^{i},N_{i}\xleftrightarrow{ 
K_{S}^{i} }S_{j},\# (N_{i}\xleftrightarrow{ K_{S}^{i} 
}S_{j}),V_{i}\rbrace \tag{13}
\end{equation}

The (13) and believe rule comprehends that
\begin{equation}
S_{j}\models BS_{j}\models (N_{i}\xleftrightarrow{ K_{S}^{i} }S_{j}) \tag{14}
\end{equation}

The logic belief proves that $S_{j}$ is confident and believes that $
K_{S}^{i}$ is issued by $BS_{j}$; moreover, the freshness of the key 
also suggests that $BS_{j}$is alive and running the protocol with $
S_{j}$ and $N_{i}$.

The (13), (14) and jurisdiction rule concludes that (15 Goal-3)

\begin{equation}
S_{j}\models (N_{i}\xleftrightarrow{ K_{S}^{i} }S_{j}) \tag{15}
\end{equation}

From I4, we have

\begin{equation}
N_{i}\prec \lbrace n_{1},,T_{R},N_{i}\xleftrightarrow{ K_{S}^{i} 
}S_{j},\# (N_{i}\xleftrightarrow{ K_{S}^{i} 
}S_{j}),N_{i}\xleftrightarrow{ n_{S}^{i} }BS_{j}\rbrace _{H(X_{S})} \tag{16}
\end{equation}

The (16), A5 and message meaning rule comprehends that
\begin{equation}
N_{i}\models BS_{j}|\sim \lbrace n_{1},,T_{R},N_{i}\xleftrightarrow{ 
K_{S}^{i} }S_{j},\# (N_{i}\xleftrightarrow{ K_{S}^{i} }S_{j})\rbrace \tag{17}
\end{equation}

The (17), A4 and freshness conjuncatenation rule infers that
\begin{equation}
N_{i}\models \# \lbrace n_{1},,T_{R},N_{i}\xleftrightarrow{ K_{S}^{i} 
}S_{j},\# (N_{i}\xleftrightarrow{ K_{S}^{i} }S_{j})\rbrace \tag{18}
\end{equation}

The (17), (18) and nonce verification rule deduce that

\begin{equation}
N_{i}\models BS_{j}\models \lbrace n_{1},,T_{R},N_{i}\xleftrightarrow{ K_{S}^{i} 
}S_{j},\# (N_{i}\xleftrightarrow{ K_{S}^{i} }S_{j})\rbrace \tag{19}
\end{equation}

The (19) and believe rule infers that
\begin{equation}
N_{i}\models BS_{j}\models \lbrace N_{i}\xleftrightarrow{ K_{S}^{i} 
}S_{j},\rbrace \tag{20}
\end{equation}

The (19), (20) and jurisdiction rule concludes that (21 Goal-1)

\begin{equation}
N_{i}\models \lbrace N_{i}\xleftrightarrow{ K_{S}^{i} }S_{j},\rbrace \tag{21}
\end{equation}

From I5, we have

\begin{equation}
N_{i}\prec IDS_{j} \tag{22}
\end{equation}

The (15), (21), (22) and meaning rule comprehends that (23 Goal-4)
\begin{equation}
S_{j}\models N_{i}\models \lbrace N_{i}\xleftrightarrow{ K_{S}^{i} 
}S_{j},\rbrace \tag{23 }
\end{equation}

From I6,we have 
\begin{equation}
S_{j}\prec n_{1} \tag{24}
\end{equation}

The (15), (21), (23) and nonce verification rule deduce that (25 Goal-2)

\begin{equation}
N_{i}\models S_{j}\models \lbrace N_{i}\xleftrightarrow{ K_{S}^{i} 
}S_{j},\rbrace \tag{25 }
\end{equation}

\begin{enumerate}
\item \textit{BAN Logic Analysis of UAAPP:}
\end{enumerate}
The UAAP protocol should achieve the following goals:

\begin{enumerate}[label=G\arabic*]
\item $U_{i}\models (U_{i}\xleftrightarrow{ K_{S}^{i} }S_{j})$
\item $BS_{j}\models U_{i}\models (U_{i}\xleftrightarrow{ K_{S}^{i} }S_{j})$
\item $U_{i}\models BS_{j}\models (U_{i}\xleftrightarrow{ K_{S}^{i} }S_{j})$
\end{enumerate}
Idealization of UAAP:

\begin{enumerate}[label=I\arabic*]
\item $U_{i}\to BS_{j}:\lbrace n_{0},(U_{i}\xleftrightarrow{ 
n_{S}^{i} }BS_{j})\rbrace _{H(X_{S})}$ 
\item $BS_{j}\to U_{i}:\lbrace n_{0},n_{1},(U_{i}\xleftrightarrow{ 
K_{S}^{i} }S_{j}, \# (U_{i}\xleftrightarrow{ K_{S}^{i} 
}S_{j}),n_{0}^{i})_{K_{l}^{i}=H(\zeta 
_{l}^{j},H(n_{0}^{i}))},(V_{i})_{K_{G}^{i}}, 
T_{R},N_{i}\xleftrightarrow{ K_{S}^{i} 
}S_{j},\# (U_{i}\xleftrightarrow{ K_{S}^{i} 
}S_{j}),U_{i}\xleftrightarrow{ n_{S}^{i} }BS_{j}\rbrace _{H(X_{S})}
$
\item $U_{i}\to BS_{j}:(n_{1})_{K_{S}^{i}}$
\end{enumerate}
Initial State Assumptions of UAAP:
\begin{enumerate}[label=A\arabic*]
\item $BS_{j}\models \# (n_{0})$
\item $U_{i}\models \# (n_{1})$
\item $N_{i}\models (U_{i}\xleftrightarrow{K_{TS}=H(X_{S}) }BS_{j})$ 
\item $BS_{j}\models (U_{i}\xleftrightarrow{K_{TS}=H(X_{S}) }BS_{j})$
\item $S_{j}\models N_{i}\models (N_{i}\xleftrightarrow{K_{TS}=H(X_{S}) }BS_{j})
$
\item $N_{i}\models BS_{j}\models (N_{i}\xleftrightarrow{K_{TS}=H(X_{S}) }BS_{j})
$
\end{enumerate}
Let us analyze the protocol to show that $UAAP $achieves the mentioned 
goals:

From I1, we have
\begin{equation}
BS_{j}\prec \lbrace n_{0},(U_{i}\xleftrightarrow{ n_{S}^{i} 
}BS_{j})\rbrace _{H(X_{S})} \tag{26}
\end{equation}

The (26), A4 and message meaning rule infers that
\begin{equation}
BS_{j}\models U_{i}|\sim \lbrace n_{0},(U_{i}\xleftrightarrow{ n_{S}^{i} 
}BS_{j})\rbrace \tag{27}
\end{equation}

The A1 and freshness conjuncatenation comprehend that 
\begin{equation}
BS_{j}\models \# \lbrace n_{0},(U_{i}\xleftrightarrow{ n_{S}^{i} 
}BS_{j})\rbrace \tag{28}
\end{equation}

The (27), (28) and nonce verification rule deduces that
\begin{equation}
BS_{j}\models \lbrace U_{i}\models n_{S}^{i}, n_{0},(U_{i}\xleftrightarrow{ 
n_{S}^{i} }BS_{j})\rbrace \tag{29}
\end{equation}

The (29) and believe rule infers that
\begin{equation}
BS_{j}\models U_{i}\models (U_{i}\xleftrightarrow{ n_{S}^{i} }BS_{j}) \tag{30}
\end{equation}

From A1, (30) and jurisdiction rule, it concludes
\begin{equation}
BS_{j}\models (U_{i}\xleftrightarrow{ n_{S}^{i} }BS_{j}) \tag{31}
\end{equation}

This belief confirms that $BS_{j}$ has received a message from a 
legitimate$N_{i}$.

From M2, we have
\begin{equation}
U_{i}\prec \lbrace n_{1},,T_{R},U_{i}\xleftrightarrow{ K_{S}^{i} 
}S_{j},\# (U_{i}\xleftrightarrow{ K_{S}^{i} 
}S_{j}),U_{i}\xleftrightarrow{ n_{S}^{i} }BS_{j}\rbrace _{H(X_{S})} \tag{32}
\end{equation}

The (32), A3 and message meaning rule comprehends that
\begin{equation}
U_{i}\models BS_{j}|\sim \lbrace n_{1},,T_{R},U_{i}\xleftrightarrow{ 
K_{S}^{i} }S_{j},\# (U_{i}\xleftrightarrow{ K_{S}^{i} }S_{j})\rbrace \tag{33}
\end{equation}

The (33), A2 and freshness conjuncatenation rule infers that
\begin{equation}
U_{i}\models \# \lbrace n_{1},,T_{R},U_{i}\xleftrightarrow{ K_{S}^{i} 
}S_{j},\# (U_{i}\xleftrightarrow{ K_{S}^{i} }S_{j})\rbrace \tag{34}
\end{equation}

The (33), (34) and nonce verification rule deduce that
\begin{equation}
U_{i}\models BS_{j}\models \lbrace n_{1},,T_{R},U_{i}\xleftrightarrow{ K_{S}^{i} 
}S_{j},\# (U_{i}\xleftrightarrow{ K_{S}^{i} }S_{j})\rbrace \tag{35}
\end{equation}

The (35) and believe rule infers that (36 Goal-2)
\begin{equation}
U_{i}\models BS_{j}\models \lbrace U_{i}\xleftrightarrow{ K_{S}^{i} 
}S_{j},\rbrace \tag{36 }
\end{equation}

The (35), (36) and jurisdiction rule concludes that (37 Goal-1)

\begin{equation}
U_{i}\models \lbrace U_{i}\xleftrightarrow{ K_{S}^{i} }S_{j}\rbrace \tag{37 }
\end{equation}

From I3, we have 
\begin{equation}
BS_{j}\prec n_{1} \tag{38}
\end{equation}

The (36), (37), (38) and meaning rule comprehends that (39 Goal-3)
\begin{equation}
BS_{j}\models U_{i}\models \lbrace U_{i}\xleftrightarrow{ K_{S}^{i} 
}S_{j},\rbrace \tag{39 }
\end{equation}

\newpage
\paperwidth=\pdfpageheight
\paperheight=\pdfpagewidth
\pdfpageheight=\paperheight
\pdfpagewidth=\paperwidth
\newgeometry{layoutwidth=297mm,layoutheight=210 mm, left=2.7cm,right=2.7cm,top=1.8cm,bottom=1.5cm, includehead,includefoot}
\fancyheadoffset[LO,RE]{0cm}
\fancyheadoffset[RO,LE]{0cm}
\begin{table}[H]
\caption{Scyther tool parameter settings.}\label{tab1}
\small 
\centering
\begin{tabular}{cc}
\toprule
{\bf Parameter} & {\bf Settings} \\\midrule
Number of Runs & 1$\sim$3 \\
Matching Type & Find all Type Flaws \\
Search pruning & Find All Attacks \\
Number of pattern per claim & 10 \\ 
\bottomrule 
\end{tabular}
\end{table}
\unskip

\begin{table}[H]
\caption{Comparison of authentication properties.}\label{tab2}
\centering
\begin{tabular}{cccccccccccccccccccccc}
\toprule
\multirow{1}{*}{{\bf Claims}} & \multicolumn{3}{C{1.5cm}}{\bf H. Tseng \cite{16}}& \multicolumn{3}{C{1.5cm}}{\bf Yoo et~al. \cite{17}}& \multicolumn{3}{C{1.5cm}}{\bf Kumar et~al. \cite{18}}& \multicolumn{3}{C{1.5cm}}{\bf Quan et~al. \cite{19}} & \multicolumn{3}{C{1.5cm}}{\bf Farash et~al. \cite{20}} & \multicolumn{3}{C{1.5cm}}{\bf Y. Lu et~al. \cite{21}} & \multicolumn{3}{C{1.5cm}}{\bf SMSN} \\

\midrule
& $U_i$ & $N_i$ & $GW$ & $U_i$ & $N_i$ & $GW$ & $U_i$ & $N_i$ & $GW$ & $U_i$ & $N_i$ & $GW$ & $U_i$ & $N_i$ & $GW$ & $U_i$ & $N_i$ & $GW$ & $U_i$ & $N_i$ & $GW$ \\

Aliveness & N & N & N & N & N & N & Y & N & N & Y & N & N & Y & Y & N & N & N & Y & Y & Y & Y \\
Weak Agreement & N & N & N & N & N & N & Y & N & N & Y & N & N & Y & Y & N & N & N & Y & O & Y & Y \\
Non-injective Agreement & N & N & N & N & N & N & O & N & N & Y & N & N & Y & Y & N & N & N & Y & O & Y & Y \\
Non-injective Synch. & N & N & N & N & N & N & O & N & N & Y & N & N & Y & Y & N & N & N & Y & O & Y & Y \\\bottomrule 
\end{tabular}
\begin{tabular}{@{}cc@{}}
\multicolumn{1}{p{\linewidth-2cm}}{\footnotesize \justifyorcenter{N = Authentication claim is not fulfilled; Y = Authentication claim is fulfilled; O = Authentication claim is fulfilled but falsified for protocol instances >3.}}
\end{tabular}
\end{table}
\newpage
\restoregeometry
\paperwidth=\pdfpageheight
\paperheight=\pdfpagewidth
\pdfpageheight=\paperheight
\pdfpagewidth=\paperwidth
\headwidth=\textwidth

\subsection{Verifying Protocol Using Scyther Tool}

The previous section proved that according to the BAN logic the SMSN is 
a secure authentication scheme. The BAN logic provided a foundation for 
the formal analysis of security protocols, but few attacks can slip 
through the BAN logic \cite{32}. For further proof of the strength of the 
SMSN protocol suite, we implemented the SMSN and \cite{16,17,18,19,20,21} schemes in 
the automated security protocol analysis tool, Scyther \cite{11,12}. Our 
proposed scheme provides a strong defense against known attacks in the 
presence of an intruder (Z) which is capable of regulating the 
communication channel, redirecting, spoofing, replaying or blocking the 
messages. It has initially known information, e.g., IDs and public keys 
of all users, and any intercepted message is additional information to 
the current information set (S), i.e., $S\cup \{m\}$. It can generate a fresh 
message from known data, e.g., $S\vdash m$, and can run multiple instances of 
the protocol. The Scyther tool verifies the protocol claims and checks 
the possibility of attacks against the protocol. The claims are the 
event that describes the design and security properties of the 
authentication protocol. We consider four claims as defined below; for a 
detailed description of protocol claims, please refer to \cite{13,14}.

In Scyther the protocol is modeled as an exchange of messages among 
different participating~'roles'; for instance, in sensor node 
authentication, the sensor node is in the role of initiator, the sink~is 
in the role of responder and the base station is in the role of a 
server. The Scyther tool integrates the authentication properties into 
the protocol specification as a claim event. We tested our protocol~\cite{16,17,18,19,20,21} employing claims, as mentioned earlier, with the parameter 
settings given in Table \ref{tab1}. 

The results are shown in Table \ref{tab2}. It is clear that in the presence of 
an intruder (as defined above), our protocol qualifies all the protocol 
claims and no attacks were found. Hence, for a large number of systems 
and scenarios, our protocol guarantees safety against a large number of 
known attacks, such as impersonating, man-in-middle and replay~attacks, 
etc. In contrast, \cite{16,17,18,19,20,21} are susceptible to several attacks and 
failed to fulfill the authentication claims. Moreover, in protocol 
schemes \cite{16,17,18,19,20,21}, there is a lack of sender-receiver binding 
verification, and an intruder can exploit this situation to impersonate 
a sensor node and run multiple instances of the protocol to launch 
multiplicity and man-in-middle~attacks.

\section{Performance Analysis }
Although the SMSN authentication protocol suite covers authentication procedures for both sensor and user nodes, for the sake of simplicity, we compare the efficiency of the SMSN with user authentication protocols. We compared the efficiency of the SMSN user authentication protocols considering computational cost, message complexity and time synchronization requirements to that of~\cite{16,17,18,19,20,21}. Unlike in \cite{16,17,18,19,20,21}, the SMSN allows the user to be 
authenticated with both sensor and sink~nodes. 

The total computation cost is estimated as the sum of the total number 
of $E$~=~encryptions/decryptions, $M$ = Multiplications, $H$ = 
hash, $X$ = XOR, and $T$ = Time Synchronization operations. 
Moreover, we assume that cryptographic hash and symmetric 
encryption/decryption operations have computational complexity similar 
to $O(m)$, where m is the size of the message. All the schemes in 
\cite{16,17,18,19,20,21} considered that the registration process took place via a 
secure channel. We assume that in all schemes the user node $U_{i} $ 
and Gateway $GW_{j}$ exchange the registration information using a 
secure encrypted channel. Furthermore, one encrypted unicast message 
requires two $E $operations, one for encryption and one for 
decryption. Similarly, a broadcast message to $N $recipient adds the $
(N+1)E $ operations in the total computational complexity.

\newpage
\paperwidth=\pdfpageheight
\paperheight=\pdfpagewidth
\pdfpageheight=\paperheight
\pdfpagewidth=\paperwidth
\newgeometry{layoutwidth=297mm,layoutheight=210 mm, left=2.7cm,right=2.7cm,top=1.8cm,bottom=1.5cm, includehead,includefoot}
\fancyheadoffset[LO,RE]{0cm}
\fancyheadoffset[RO,LE]{0cm}

\begin{table}[H]
\caption{Performance comparison of SMSN with well-known user authentication protocols.}\label{tab3}
\small 
\centering
\begin{tabular}{cccccc}
\toprule
{\bf Schemes} & {\bf Phase} & {\bf Comp. Complexity} & {\bf Comm. Complexity} & {\bf Comm. Cost in Bytes} & {\bf Time Synch.} \\
\midrule
\multirow{2}{*}{H. Tseng\cite{16}} & Registration & $(4+N)E+1H$ & $2UC+1BC $ & $(2N+3)int$ & - \\
& Login-Authentication & $6H+1X$ & $4UC$ & $2CK+7int$ & $1T$ \\
\multirow{2}{*}{Yoo~et~al.~\cite{17}} & Registration & $(4+N)E+5H+2X$ & $2UC+1BC$ & $(5+N)CK+1int$ & - \\
& Login-Authentication & $19H+2X$ & $6UC$ & $5CK+8int$ & $1T$ \\
\multirow{2}{*}{Kumar~et~al.~\cite{18}} & Registration & $4H+3X $ & $2UC$ & $5CK+3int$ & - \\
& Login-Authentication & $14H+2X$ &$3UC$ & $6CK+11int$ & $2T$ \\
\multirow{2}{*}{Quan~et~al.~\cite{19}} & Registration & $12E+9H+1X+7M$ & $4UC$ & $4CK+9int+4f$ & $4T$ \\
& Login-Authentication & $18H+2X$ & $4UC$ & $8CK+16int$ & $4T$ \\
\multirow{2}{*}{Farash~et~al.~\cite{20}} & Registration & $4E+6H+2X$ & $2UC$ & $4CK+1int$ & $2T$ \\
& Login-Authentication & $30H+16X$ & $4UC$ & $17CK+5int$ & $4T$ \\
\multirow{2}{*}{Y. Lu~et~al.~\cite{21}} & Registration & $4E+10H+2X$ & $2UC$ & $4CK+1int$ & - \\
& Login-Authentication & $8E+17H+15X$ & $4UC$ & $4CK+18int$ & $3T$ \\
\multirow{2}{*}{SMSN (User-Sink)} & Registration & $8E+5H$ & $3UC$ & $3CK+10int$ & - \\
& Login-Authentication & $8E+2H$ & $3UC$ & $3CK+8int$ & $Optional$ \\
\multirow{2}{*}{SMSN (User-Sensor)} & Registration & $8E+5H$ & $3UC$ & $3CK+10int$ & - \\
& Login-Authentication & $9E+2H$ & $4UC$ & $5CK+12int$ & $Optional$ \\
\bottomrule 
\end{tabular}
\begin{tabular}{@{}cc@{}}
\multicolumn{1}{p{\linewidth-2cm}}{\footnotesize \justifyorcenter{{\it Computational Complexity:} $E $ = encryptions/decryptions, $ Ex$ = modular exponentiations, $M$ = multiplications, $ H$ = hash operation, $ X $ = $XOR$ operation , $N$ = Total number of nodes , $T$~=~Time Synchronization operation; {\it Communication Complexity and Cost:} $BC$ = Broadcast Message, $UC$ = Unicast Message, $ CK$ = 32 bytes (Represents Size of Cryptographic KeyHash and Signature), $Int$ = 4 bytes ( Represents Size of nonce, Node Ids and Integers), $f$ = 8 bytes (Represents Size of the floating point real number.}}
\end{tabular}
\end{table}
\newpage
\restoregeometry
\paperwidth=\pdfpageheight
\paperheight=\pdfpagewidth
\pdfpageheight=\paperheight
\pdfpagewidth=\paperwidth
\headwidth=\textwidth

From Table \ref{tab3} we can see that computational complexity of SMSN 
authentication protocols in the~registration phase is slightly more 
expensive compared to Kumar~et~al.~\cite{18} and Farash~et~al.~\cite{20}; 
however; in the authentication phase, the SMSN authentication protocols 
completely outperform~Kumar~et~al.~\cite{18} and Farash~et~al.~\cite{20}. 
H. Tseng \cite{16} and Yoo~et~al.~\cite{17} are the most computationally 
expensive schemes during the registration phase, but in the 
authentication phase, H. Tseng \cite{16} is the most efficient scheme 
followed by the SMSN. However, regarding sensor node computational 
efficiency, in our scheme, the overall workload of a sensor node is very 
low. Moreover, unlike the~\cite{15,17,18,19} schemes, the SMSN does not require 
time synchronization between the Gateway and user node.

\begin{figure}[H]
\centering
\includegraphics[width=8 cm]{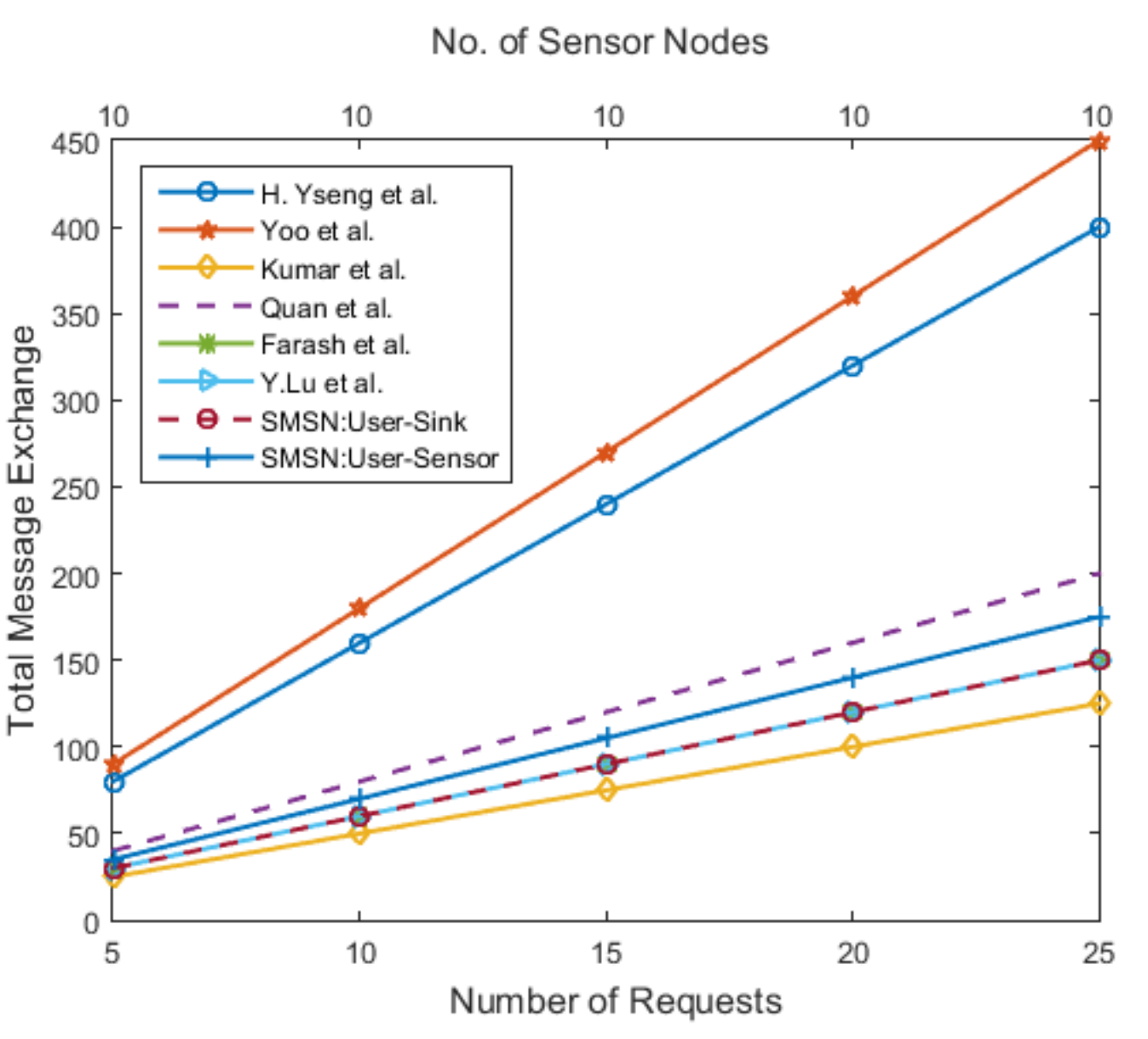}
\caption{The total number of message exchanged between constant number of sensor nodes with different number of new users' requests. }\label{fig13}
\end{figure}

\begin{figure}[H]
\centering
\includegraphics[width=8 cm]{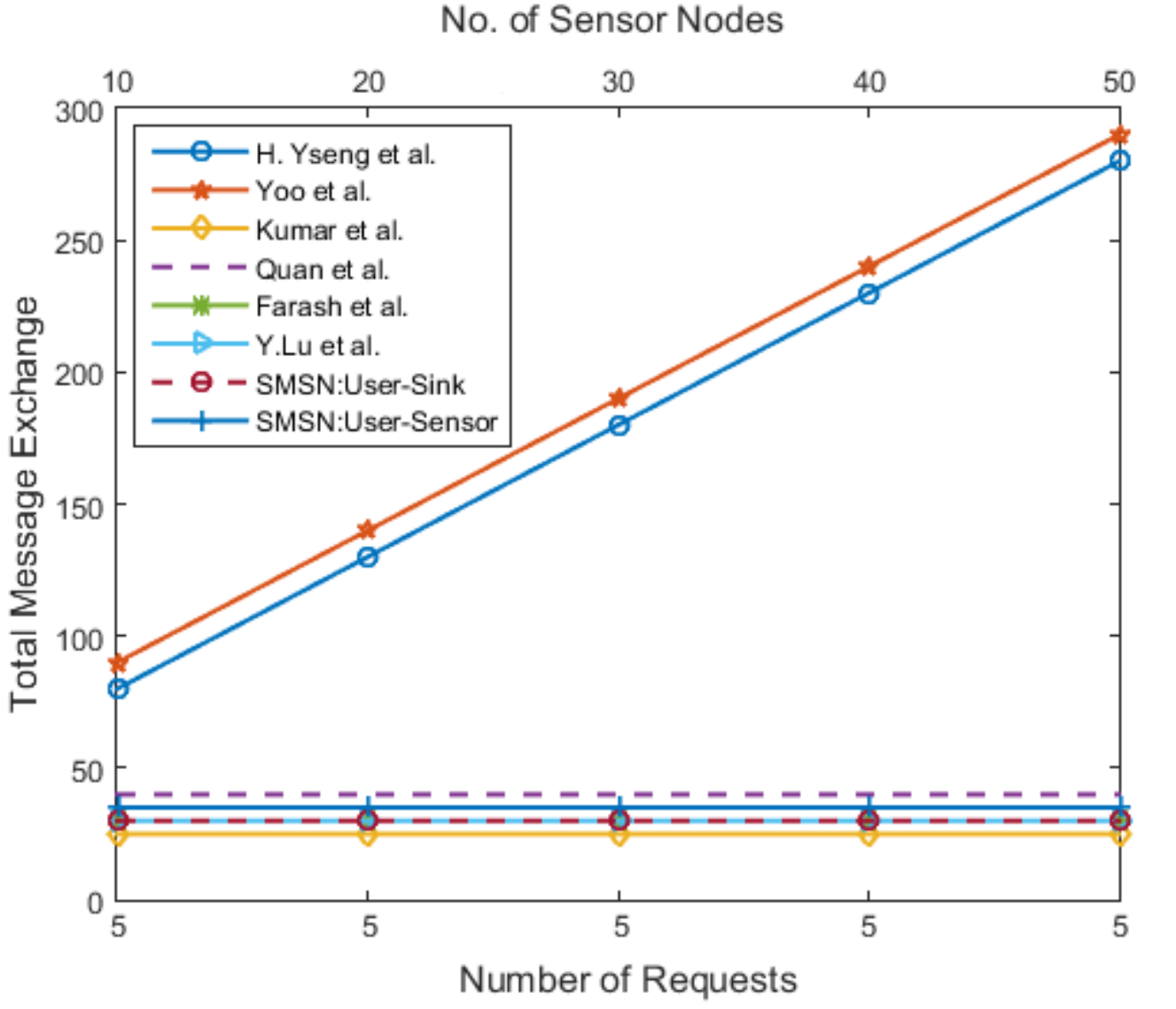}
\caption{The total number of message exchanged for constant number of new users request in different network size. }\label{fig14}
\end{figure}

\begin{figure}[H]
\centering
\includegraphics[width=8 cm]{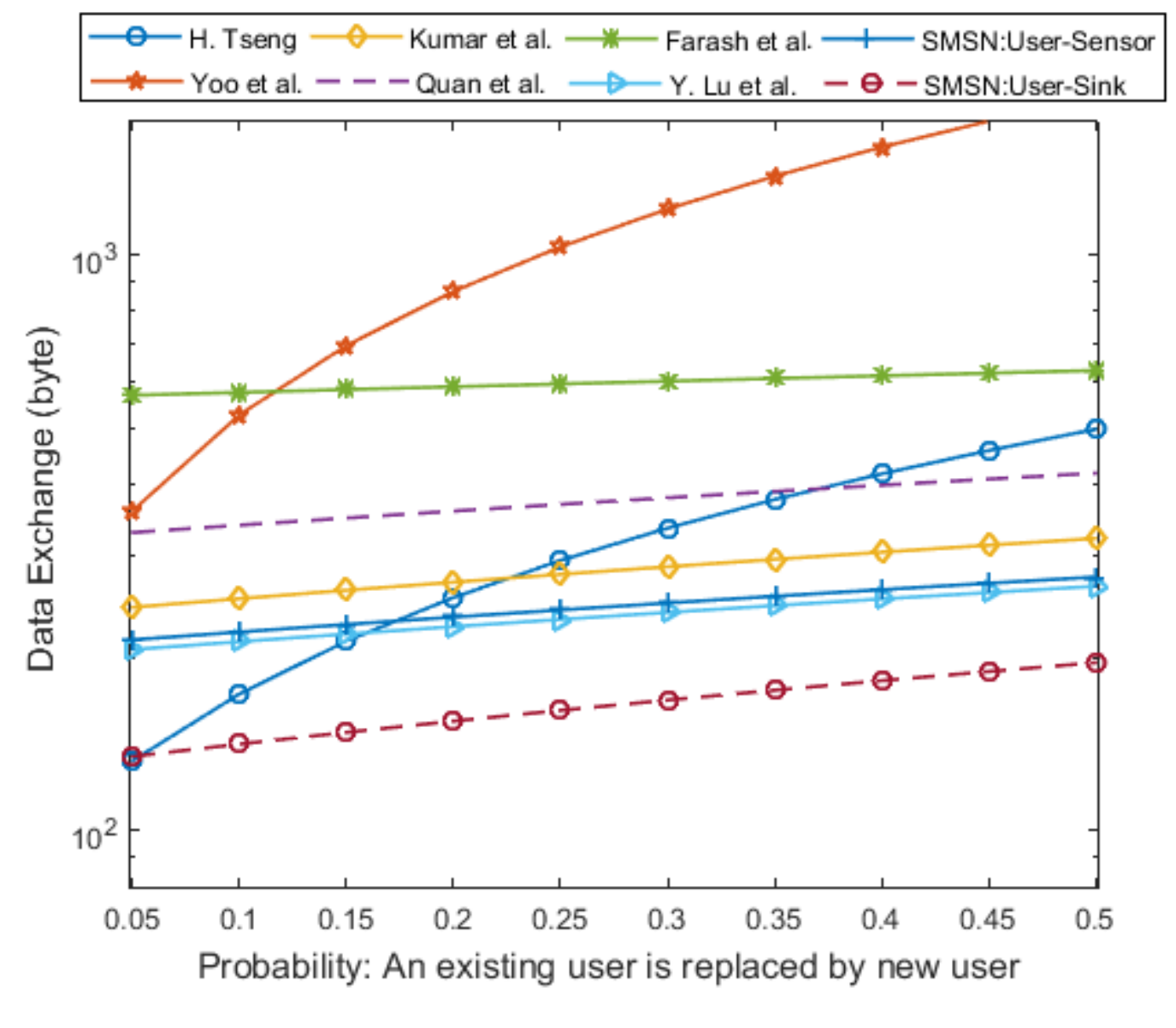}
\caption{The amount of data exchanged for less to highly dynamic sensor network. }\label{fig15}
\end{figure}

The communication complexity is calculated as the sum of the total 
unicast and broadcast message exchange. Figure \ref{fig13}
shows the overall 
communication complexity in a WSN when the number of sensor nodes is 
fixed in the network, and the number of new users' requests is 
constantly increasing. The message complexity of \cite{16} and \cite{17} 
increases multiplicatively by increasing the number of nodes; 
conversely, it grows slowly in the case of the SMSN, \cite{18}, \cite{19}, 
and \cite{20}. Figure~\ref{fig14}
shows the overall communication complexity for 
various network sizes with the same number of user requests. This~metric 
is only useful for a WSN with mobile sensor and user nodes. The message 
complexity of \cite{16,17} increases rapidly with an increase in the 
number of new users' requests; conversely, it remains constant with the 
SMSN, \cite{18}, \cite{19}, and \cite{20}. This~suggests that the schemes 
proposed in \cite{16,17} are not suitable for highly dynamic mobile WSNs 
where the frequency of leaving and joining the network is high. 
However, a more interesting comparison in terms of communication 
efficiency is the comparison based on the amount of data exchanged 
during the protocol run. The numerical results are taken for a dynamic 
and mobile sensor network consisting of 100 nodes. The probability that 
a new user may join the network and an existing user may leave the 
system defines how frequently the users join and leave the network. The 
average communication cost per user is calculated for the dynamic 
probability of 0.05 to 0.5, and the results are shown in Figure \ref{fig15}.

SMSN user-sink authentication outperforms all other schemes from less dynamic 
to highly dynamic networks. However, in a less dynamic network with a 
dynamic probability of less than 0.05, the SMSN user-sensor 
authentication is slightly more expensive than the scheme of H. Tseng~\cite{16}. Even though in a less dynamic system the SMSN user-sensor 
authentication is slightly more expensive than~\cite{16}, the performance 
gap decreases, and for highly mobile and dynamic networks, SMSN performs 
better than the scheme of H. Tseng~\cite{16}.
\unskip
\section{Conclusions }
Due to the recent growth in WSN technologies, we have observed an 
enormous paradigm shift in sensor network applications. The 
authentication and security goals of a sensor network have become more 
crucial and challenging. Most of the user and sensor node authentication 
schemes for WSNs have been developed without taking into account the 
requirements of integrating WSNs with emerging technologies such as IoT. 
We developed an SMSN scheme considering the requirements of mobile and 
dynamic WSN applications as discussed in Section-III. We noted that the 
user authentication schemes designed for sensor networks \cite{16,17,18,19,20,21} do 
not meet the authentication properties; for example, the execution of 
these schemes in the Scyther tool revealed that the participating 
entities failed to achieve wider objectives (defined as protocol claims 
in Section 5-C) of the authentication protocol. Finally, we compared the 
efficiency of the SMSN user authentication protocols with the schemes in 
\cite{16,17,18,19,20,21}. We~observed that concerning the computational cost, our 
scheme is slightly more expensive compared to \cite{18,20} during the 
registration phase but the SMSN totally outperforms both in the 
authentication phase. Regarding message complexity our proposed scheme 
totally outperforms~\cite{16,17,18,19,20}; however, the performance of the scheme 
of Y. Lu~et~al.~\cite{21} is close to the SMSN. Finally, unlike the 
schemes in~\cite{16,17,18,19,20,21}, the SMSN does not require time synchronization 
between the Gateway (base station) and the user node. The main focus of 
this work was to discuss and provide solutions for the emerging 
challenge that has emerged from the integration of the WSN in IoT 
applications. To prove the usability of the proposed scheme, we made a 
comprehensive security and performance analysis and simulated the 
proposed idea in an automated protocol verifier tool, the Scyther. 
However, for future work, it will be interesting to investigate the 
usability of the SMSN by implementing it on an application specific 
testbed. Moreover, in the near future it will be possible to incorporate 
the basic Internet functionality in the sensor node. We believe it will 
further enhance the application scenarios for the SMSN; for instance a 
user device will be able to collect real-time sensor data remotely via 
the Internet. Moreover, we are further investigating the usage of SMSN for the promising future internet architecture, known as Name-Data-Networking (NDN), which is extensively studied in the literature~\cite{55,56,57,58,59,60}. In NDN the contents verification is achieved by the use of asymmetric cryptography. We~argue that in future especially in IoT application scenarios the devices will be resource constraint devices; for instance, the sensor network is going to be the part of IoT. For resource constraint devices the asymmetric cryptography is computationally expensive. We believe that SMSN protocol suit with some modifications can be a suitable candidate for NDN internet architecture.
\vspace{6pt}


\acknowledgments{ This work was supported by the Ministry of Trade, Industry \& Energy (MOTIE, Korea) under Industrial Technology Innovation Program. No. 10065742, Standardization of Vertical and Horizontal Smart-factory Integration.}

\authorcontributions{Muhammad Bilal proposed the scheme, and performed the security and performance analysis. He also wrote the experiment code, conducted the experiments and evaluated the performance of the proposed scheme. Shin-Gak Kang amended the manuscript.}


\abbreviations{The following abbreviations are used in this manuscript:\\

\noindent 
\begin{tabular}{ll}
WSN&Wireless Sensor Network\\
IoT&Internet of Things\\
SMSN-Protocol&Secure Mobile Sensor Network Protocol  \\
BAN-logic&Burrows Abadi Needham logic \\
SAAP&Sensor Activation and Authentication Protocol \\
SRP1&Sensor Re-Authentication Protocol-1 \\ 
SRP2&Sensor Re-Authentication Protocol-2 \\
UAAP&User Activation and Authentication Protocol \\ 
USiAP&User-Sink Authentication Protocol   \\ 
USeAP&User- Sensor Authentication Protocol    
\end{tabular}
}


\bibliographystyle{mdpi}

\renewcommand\bibname{References}


\section*{Authors' Biographies}

\setlength\intextsep{0pt}
\begin{wrapfigure}{l}{0.2\textwidth}
\includegraphics[width=0.2\textwidth]{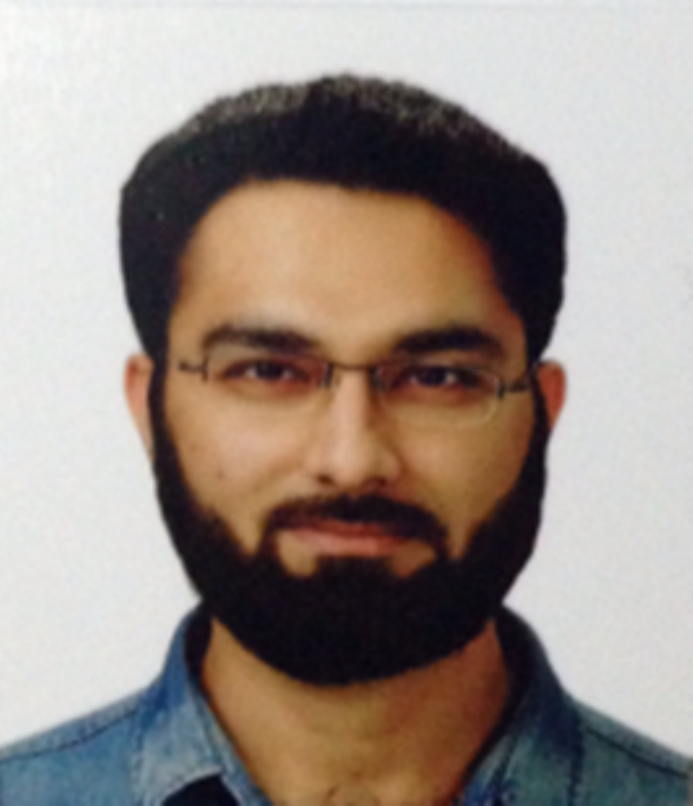}
\end{wrapfigure}
\textbf{Muhammad Bilal} has received his BS degree in computer systems engineering from University of Engineering and Technology, Peshawar, Pakistan and MS in computer engineering from Chosun University, Gwangju, Rep. of Korea. Currently, he is Ph.D. student at University of Science and Technology, Korea at Electronics and Telecommunication Research Institute Campus, Daejeon, Rep. of Korea. He has served as a reviewer of various international Journals including IEEE Transactions on Network and Service Management, IEEE Access, IEEE Communications Letters, Journal of Network and Computer Applications, Personal and Ubiquitous Computing and International Journal of Communication Systems. He has also served as a program committee member on many international. His primary research interests are Design and Analysis of Network Protocols, Network Architecture, and Future Internet.  

\setlength\intextsep{0pt}
\begin{wrapfigure}{l}{0.2\textwidth}
\includegraphics[width=0.2\textwidth]{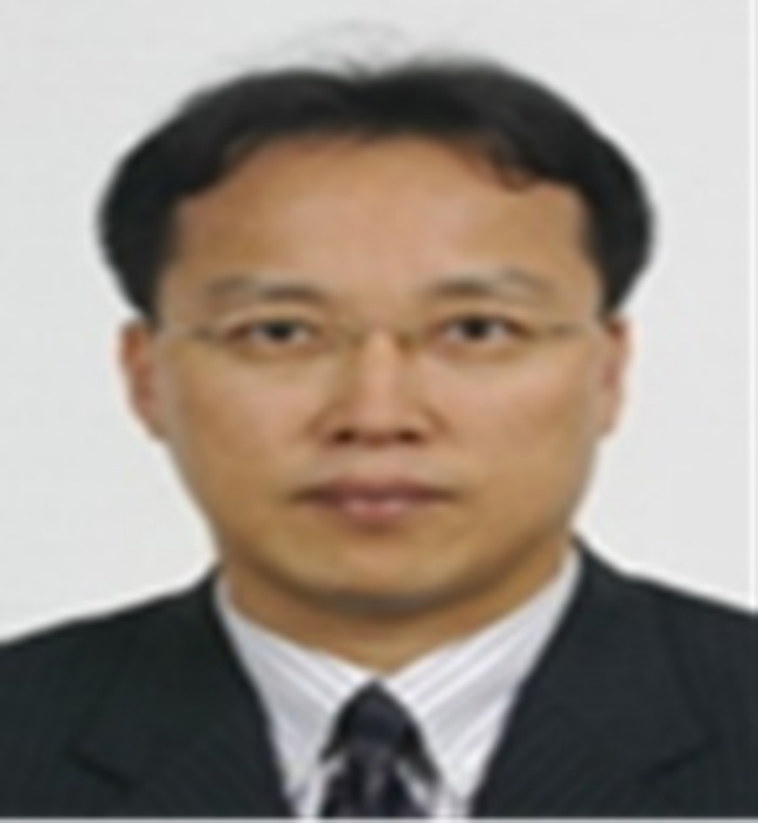}
\end{wrapfigure}
\medskip
\textbf{Shin-Gak Kang} received his BS and MS degree in electronics engineering from Chungnam National University, Rep. of Korea, in 1984 and 1987, respectively and his Ph.D. degree in information communication engineering from Chungnam National University, Rep of Korea in 1998. Since 1984, he is working with Electronics and Telecommunications Research Institute, Daejeon, Rep. of Korea, where he is a principal researcher of infrastructure standard research section. From 200 to 2008 he served as an editor with ETRI Journal. From 2008 he is a professor at the Department of Information and Communication Network Technology, University of Science and Technology, Korea. He is actively participating in various international standard bodies as a Vice-chairman of ITU-T SG11, Convenor of JTC 1/SC 6/WG 7, etc. His research interests include multimedia communications and Applications, ICT converged services, contents networking, and Future Network.

\end{document}